\documentclass[fleqn,usenatbib,useAMS]{mnras}

\usepackage{newtxtext,newtxmath,lastpage}

\usepackage[T1]{fontenc}

\DeclareRobustCommand{\VAN}[3]{#2}
\let\VANthebibliography\thebibliography
\def\thebibliography{\DeclareRobustCommand{\VAN}[3]{##3}\VANthebibliography}

\usepackage{graphicx}	

\usepackage{amsmath}	
\usepackage{bm}

\title[Revealing the disc structure in CS-AGN]{Revealing the accretion disc structure in Changing State AGN: NGC1566 }

\author[H. Wang, C. Done, R. Middei\& X.-B. Wu]{
Huimei Wang,$^{1,2,4}$
Chris Done,$^{2}$\thanks{E-mail: chris.done@durham.ac.uk}
Riccardo Middei,$^{3}$
Xue-Bing Wu$^{1,4}$
\\
$^{1}$Department of Astronomy, Peking University, Beijing 100871, People's Republic of China\\
$^{2}$Centre for Extragalactic Astronomy, Department of Physics, University of Durham, South Road, Durham, DH1 3LE, UK\\
$^{3}$INAF Osservatorio Astronomico di Roma, Via Frascati 33, 00078 Monte Porzio Catone, Italy\\
$^{4}$Kavli Institute for Astronomy and Astrophysics, Peking University, Beijing 100871, People's Republic of China
}

\date{Accepted XXX. Received YYY; in original form ZZZ}

\pubyear{2015}

\begin{document}
\label{firstpage}
\pagerange{\pageref{firstpage}--\pageref{lastpage}}
\maketitle

\begin{abstract}

NGC1566 is a low mass ($\sim 6\times 10^6M_\odot$) 
Active Galactic Nuclei (AGN) with unusual large amplitude variabilty in its accretion flow. In the optical it cycles between 
recurring short bright episodes with strong blue/UV continuum and clear broad line emission, and a much weaker persistent level without strong broad lines, making it a repeating changing-state (CS) AGN. We re-analyse the intensive optical/UV and X-ray monitoring during the 2018 bright flare, showing for the first time the corresponding evolution of the (host galaxy subtracted) optical/UV AGN spectrum as well as the X-ray emission. As seen in other CS-AGN, the strongest spectral evolution is in a UV/soft X-ray component which likely disappears below $0.01L_{\rm Edd}$ while the hot corona and a redder optical/UV component show weaker variability. We show that the red optical emission is likely a composite of the intrinsic and reprocessed emission from the outer disc, together with diffuse continuum from the BLR, and discuss the different signatures predicted by this for continuum reverberation across the changing-state event. The historical lightcurves show
repeated changing-state events tied to unusually large amplitude discrete flares in the lightcurve. While these do not appear to be periodic, the behaviour does support previous suggestions that the accretion flow is repeatedly perturbed, perhaps by an intermediate mass black hole companion.

\end{abstract}


\section{Introduction}

Changing-State Active Galactic Nuclei (CS-AGN) give a unique insight into the time dependent evolution of the accretion flow. These objects were first found by repeated optical spectroscopy showing the appearance/disappearance of the broad line region (BLR) emission correlated with a change in the optical/UV continuum strength \citep{LaMassa2015, MacLeod2016, Yang2018, Ruan2019, MacLeod2019, Sheng2019}. These were first termed Changing-Look AGN, associated with changing X-ray  obscuration along the line of sight \citep{Risaliti2002}, but it is now clear from the correlated change in the infrared echo of the reprocessed UV illumination of the torus \citep{Sheng2017}, that the majority of these represent intrinsic changes in the accretion flow. These are better described as Changing-State (CS) AGN, and they pick out a critical luminosity of $L\sim 0.01L_{\mathrm{Edd}}$ \citep{Noda2018,Ruan2019,Guo2025} for the appearance/disapearance of the BLR. This is similar to the critical luminosity for accretion-state transitions seen in stellar-mass black hole binaries, in which the inner regions of the standard disc transition into a hot, geometrically thick and radiatively inefficient accretion flow \citep{Narayan1995}.  

However, in broad line AGN it is clear that the standard disc plus X-ray corona is not a good descriptor of the spectral energy distribution (SED). There is an  UV downturn at lower energies than predicted by the expected disc peak temperature, and this appears to match to an upturn in soft X-rays (soft X-ray excess) over that expected from a coronal power law. This is now generally interpreted in models where the inner accretion flow contains an optically thick, warm Comptonising region, located between the outer standard disc and the compact hot corona \citep{Mehdipour2011, Done2012, Kubota2018, Petrucci2018, Petrucci2020}. Broadband UV–X-ray spectroscopy is therefore essential for separating these components and for tracing how the accretion flow responds to changes in accretion rate.

So far, only a small number of CS-AGNs have sufficiently well-sampled broadband UV–X-ray observations to follow these structural changes directly. In Mrk1018 ($\sim 7\times 10^7 M_{\odot}$), the disappearance of the soft-excess component as the Eddington ratio declined was the key to interpreting the changing-look event as evidence for a changing accretion state \citep{Noda2018}. Similarly, 
ESO511-G030 ($\sim 1.7\times 10^7 M_{\odot}$; \citealt{Middei2023, Middei2026}) and 
Mkn~590 ($M_{\rm BH}\sim4.7\times10^{7}\,M_{\odot}$; \citealt{Palit2025}) have
both recently undergone 
multiwavelength rebrightening, from a dim state with little broad line emission or UV flux, to a much brighter state with strong UV and soft X-ray flux and renewed broad-line emission, suggesting the recovery of an optically thick accretion flow.

These individual AGN all show that the soft X-ray excess rises along with the UV bright component, rather than tracking the coronal X-rays which are always present at some level. This strongly favours models where the majority of the soft X-ray excess is part of the disc, rather than models connecting it to the X-ray corona by extreme relativisitically smeared reflection
\citep{Crummy2006}. 

This spectral transition is also seen in the general AGN population. At a fixed black hole mass ($\log M/M_\odot =8.0-8.5$), the UV and soft X-ray bright disc component systematically weakens relative to the (approximately constant) X-ray coronal power as the Eddington rate declines. This behaviour is equivalent to the 
non-linear correlation seen between the monochromatic UV (2500\AA) and X-ray (2keV) luminosities seen across the bright quasar population \citep{Lusso2016}.
The UV component then disappears 
below $L/L_{\rm Edd}\sim0.01$, while the coronal X-rays remain \citep{Ho2008, Hagen2024b, Kang2025}. 

All this demonstrates the power of broadband optical/UV/X-ray SED studies to
give more insight into the accretion flow structures than 
X-ray studies alone \citep{Chen2025,Jana2026}.
Here we extend the broadband CS studies to a lower mass AGN, NGC1566, with $M_{\rm BH} \sim 6\times 10^6M_\odot$. We 
use Swift/UVOT and XRT monitoring, together with an XMM-Newton snapshot, to trace the optical-UV–X-ray SED evolution of NGC1566 through its CS transition. This was triggered by an X-ray bright peak in 2018, followed by fast (month-long) decline to a dim state, with late time smaller reflares. We show that its accretion-flow evolution during the bright to dim state transition on the first decline
closely resembles that seen in higher-mass AGNs, extending the evidence for disc/warm-corona structural changes across a wider range of SMBH mass. 

In particular, the optical/UV data give the best view so far of the changing {\em shape} (as well as normalisation) of the AGN emission in this band. Even after careful host galaxy subtraction, the variable optical flux drops by much less than the UV, directly showing how the AGN optical/UV component reddens across the transition. We discuss the origin of this optical emission, with potential contributions from intrinsic and reprocessed X-ray emission from the outer disc emission plus reprocessed UV emission from the 
BLR. We use this to predict
how results from intensive broadband continuum reverberation monitoring campaigns might change across the CS events. 

We put the CS event in context of the historical behaviour of NGC1566, where repeated CS episodes are connected to unusually large amplitude discrete flare events. While these do not appear to be periodic, their nature does support models where the accretion flow is repeatedly perturbed by a binary companion, perhaps an intermediate black hole \citep{Dodd2025}. We predict that the source will undergo another CS event if the ASAS-SN \citep{Hart2023} g-band magnitude crosses 12.2~mag, and strongly urge
intensive broadband observations when this next occurs. 

\section{Data Reduction}
\label{Sec: data}

\subsection{\textit{Swift} XRT }
We used all available \textit{Swift}/XRT observations, taken in both photon-counting and window-timing modes. 
We first used the official online products provided by the UK \textit{Swift} Science Data Centre (UKSSDC) \citep{Evans2007, Evans2009} to make the \textit{Swift}/XRT count rate light curve (see upper panel of Fig.\ref{fig:swift_xrt_uvot_lc}). 

We then processed all the datasets using the standard \textit{Swift} analysis pipeline with HEASOFT v6.35.2
and CALDB version 2025-06-09
Source and background spectra were extracted using \texttt{xrtproducts} described in \citet{Evans2009}
\footnote{https://swift.gsfc.nasa.gov/analysis/xrt\_swguide\_v1\_2.pdf}. 
The source and background regions used to extract the spectra were selected by taking into account the possible pile-up effect in each individual observation. We
followed the prescriptions adopted in \citet{Evans2009} and \citet{Middei2022}. 
This uses the estimated XRT count rate from the cleaned event file to select an
aperture in order to 
optimize the signal-to-noise ratio in dim spectra and correct for pileup in bright spectra (see Table~\ref{tab:xrt_regions}). The changing extraction area corrected by running \texttt{xrtmkarf} using the point spread function. 

The background was extracted from a concentric annulus with inner and outer radii of 60 and 110 pixels, respectively \citep{Evans2009}.

The extracted source spectra were then grouped with \texttt{grppha}, requiring a minimum of 25 counts per bin in order to apply $\chi^2$ statistics in the spectral fitting. 

To improve the signal noise ratio (SNR) of the Swift/XRT spectra, we adopted a flux- and time-based grouping strategy. Individual spectra with more than 15 spectral bins after grouping were treated as sufficiently well constrained and were fitted individually. For spectra with fewer than 15 bins, we searched for observations close in time and with similar flux levels, and co-added them into a single spectrum using \texttt{ftaddspec}. 
If there were no nearby observations with comparable X-ray fluxes available we discarded the spectra if this was below 10 bins, or fitted it individually for 10-15 bins. This procedure balances keeping high temporal resolution with having 
adequate SNR for spectral analysis without combining observations from substantially different source states. 

\subsection{\textit{Swift} UVOT }

\begin{figure*}
    \centering
    \includegraphics[width=0.8\textwidth]{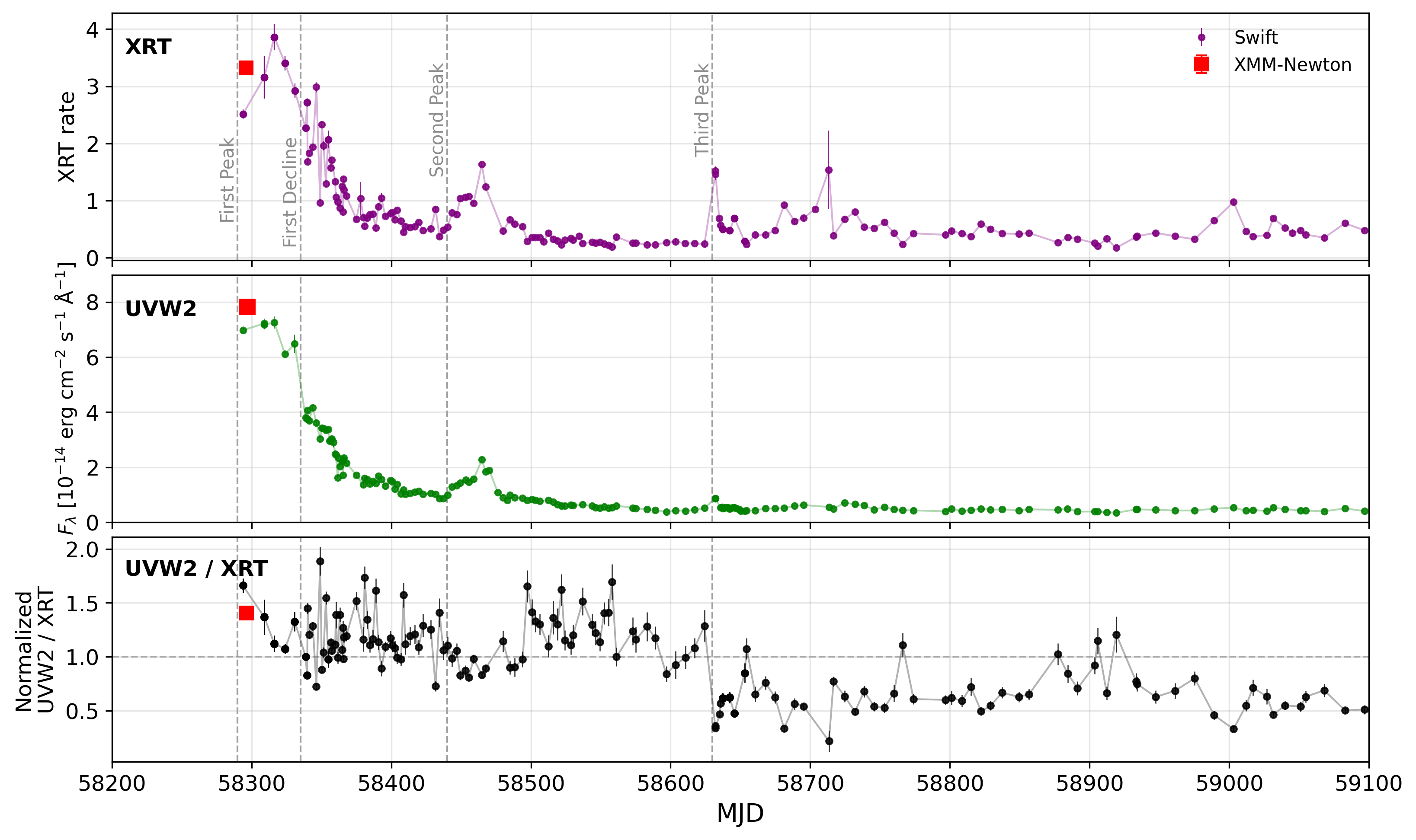}
    \caption{
    Swift/XRT and UVOT light curves of NGC~1566, together with the normalized UVW2/XRT ratio. 
    The top panel shows the Swift/XRT count-rate light curve, and the middle panel shows the UVOT flux light curve in the UVW2 filter. 
    The bottom panel presents the normalized UVW2/XRT ratio, tracing the relative evolution of the UV and X-ray emission. The red squares indicate the XMM-Newton observation, including the XMM-OM UVW2 flux and the corresponding ratio used for comparison.
    }
    \label{fig:swift_xrt_uvot_lc}
\end{figure*}

\textit{Swift}/UVOT provides observations in six filters, including three optical bands (V, B, and U) and three ultraviolet bands (UVW1, UVM2, and UVW2). We reduced the level 2 image files and performed aperture photometry using \texttt{uvotimsum} and \texttt{uvotsource}. A circular source aperture with a radius of $5^{\prime\prime}$ was adopted, centered on the target, together with a background region of an annulus between radius $27.5^{\prime\prime}$ and $35^{\prime\prime}$.

The UVW2 lightcurve is shown together with the simultaneous XRT light curve and the normalized UVW2/XRT ratio in Figure~\ref{fig:swift_xrt_uvot_lc}.
It is clear that the outburst is characterised by a fast rise in both X-ray and UVW2, folllowed by a decline with several smaller outbursts superposed. Clearly there are differences in the UV/X-ray ratio, showing that the spectral shape is evolving. The ratio is high 
during the first peak and decline, then the second peak has a stronger change in X-ray flux than UV, but during the decline from this second peak their ratio comes back to the same UV/X-ray ratio as in the first peak decline. The third peak again has a much stronger change in X-ray flux than UV, but this time the ratio does not recover during the decline, but stays low. 

The UVOT photometric measurements were converted to flux densities using the standard UVOT calibration \citep{Poole2008}. 
After subtracting the host-galaxy contribution, as described in Section~\ref{sec:host}, we constructed the UV/optical SEDs for each epoch. 
The host-subtracted UVOT fluxes were subsequently converted into PHA spectra using \texttt{ftflx2xsp}, enabling the UVOT and XRT data to be fitted jointly in \texttt{XSPEC} within a consistent spectral-fitting framework.

\subsection{XMM-Newton: EPIC PN and OM}

There is very rapid X-ray spectral evolution during the early bright phase, so we also extract the XMM-Newton data obtained close to the start of the outburst, on MJD58296.0. We extract the PN 0.3-10~keV spectra, background and arf from the XMM-Newton standard pipeline products
\footnote{
https://heasarc.gsfc.nasa.gov/FTP/xmm/data/rev0/0800840201/PPS/}
and combine with the appropriate canned response matrix.
\footnote{https://xmm-tools.cosmos.esa.int/external/xmm\_ccf/ccf/extras/responses/PN/}

In order to make a direct comparison to the \textit{Swift} data in Fig.\ref{fig:swift_xrt_uvot_lc}, we also 
convert the PN spectrum to an equivalent \textit{Swift}/XRT count rate. We fit the PN spectrum in {\sc xspec}, then use the {\sc fakeit} command to simulate this model spectrum through the XRT response. We also take the XMM-Newton OM UVW2 flux from the standard pipeline producs. The red squares in 
Fig.\ref{fig:swift_xrt_uvot_lc} show these are similar to (but slightly higher than) the 
contemporaneous \textit{Swift} fluxes (see also \citealt{Parker2019}). 

\section{Host galaxy subtraction}\label{sec:host}

\begin{figure*}
    \centering
    \includegraphics[width=0.9\textwidth]{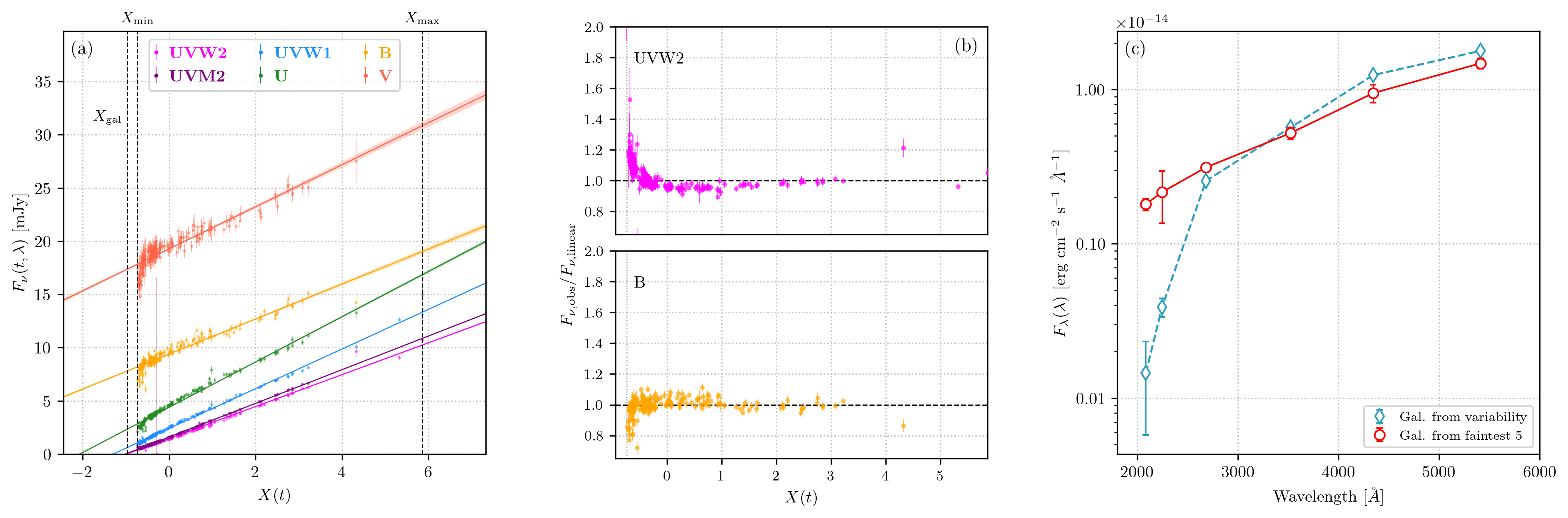}
    \caption{
    Illustration of the variability-based decomposition used to estimate the host-galaxy contribution. 
    Left: observed UVOT flux densities in different bands as a function of the normalized light-curve shape \(X(t)\). 
    The solid lines show the best-fitting linear relations for each UVOT band. 
    The vertical dashed lines mark the extrapolated host-dominated level \(X_{\rm gal}\), the minimum observed state \(X_{\min}\), and the maximum observed state \(X_{\max}\). 
    Right: comparison between the two empirical estimates of the host-galaxy contribution in the six UVOT bands. decomposition of the UV/optical SED into the mean spectrum, variable component, host-galaxy component, and AGN component.
}
    \label{fig:variability_decomposition}
\end{figure*}
The host-galaxy of NGC1566 
makes a very significant contribution to the UV/optical fluxes, especially when the AGN dims.
Host galaxy templates are a good solution where the
the UVOT aperture includes most of the host, 
but NGC1566 is so close that the standard UVOT 5" source radius encloses only a small region around the nucleus rather than the entire barred spiral (size $>4$'), making templates unreliable. 

Instead, we tried direct decomposition of high resolution HST UV imaging, which shown in Figure~\ref{fig:galfit_result}. However the nuclear region is very complex, with non-axisymmetric structures such as a 1" offset starformation region to the SW \citep{daSilva2017}
which means that a simple two-dimensional image decomposition with \texttt{GALFIT} \citep{Peng2002} does not sufficiently follow the complexity (see Fig.~\ref{fig:galfit_result}). 

Instead we estimate the host-galaxy contribution using the 
lightcurves to decompose the flux in each UVOT filter into a constant plus variable component \citep{Cackett2007}, following the 
the flux--flux analysis procedure described in Appendix~A of \citet{Kynoch2026}. 
In this method, the observed multiband flux density is described as the sum of a constant component and a variable component,
\begin{equation}
    F_{\nu}(\lambda,t) = A_{\nu}(\lambda) + S_{\nu}(\lambda) X(t),
\end{equation}
where \(X(t)\) is a dimensionless light curve normalized to have zero mean and unit variance, \(A_{\nu}(\lambda)\) represents the mean spectrum, and \(S_{\nu}(\lambda)\) represents the rms spectrum of the variable component. 
The light-curve shape \(X(t)\) is first estimated from the normalized deviations of the multiband light curves and then iteratively refined together with \(A_{\nu}(\lambda)\) and \(S_{\nu}(\lambda)\) through linear regression. 
We use all the UVOT data available in the archive for this source (see Fig.\ref{fig:ngc1566_longterm_history}), not just that from the well monitored period in 2018-2019 shown in Fig.~\ref{fig:swift_xrt_uvot_lc}. 

Once the linear flux--flux relations are obtained, they can be extrapolated to the flux level at which the variable AGN component becomes negligible, giving an estimate for the host galaxy flux in each filter (see Figure~\ref{fig:variability_decomposition}a). It is clear that the host-dominated level \(X_{\rm gal}\) lies very close to the minimum observed state \(X_{\min}\), showing that the  
faintest observed UVOT epochs of NGC1566 are strongly dominated by the host-galaxy emission. 

However, this approach clearly overestimates the host 
flux at the lowest luminosities in the V and B bands. 
Applying this decomposition to subtract the host contribution leads to unphysical negative AGN fluxes. A closer inspection reveals the problem is more widespread. The UVOT fluxes are not linearly related to the combined lightcurve, $X(t)$ (see Figure~\ref{fig:variability_decomposition}b and Appendix). A linear relation systematically overestimates the V and B band fluxes at the lowest luminosities, but there is also a systematic underestimation at slightly higher fluxes, while the UVW2 band is systematically underestimated at the lowest luminosities, but overestimated at slightly higher fluxes. 
This is showing that the {\it shape} of the optical/UV variable AGN component is changing, so this linear analysis is not accurate in detail. 

Instead, we get a  conservative estimate of the host-galaxy contribution using the 
faintest UV states observed by \textit{Swift}/UVOT. 
Among the six UVOT bands, UVW2 is the most sensitive to the AGN emission and is expected to have the largest fractional contribution from the variable accretion-disk component. 
We selected the five observations with the lowest UVW2 fluxes from the full \textit{Swift} data set, assuming that these epochs correspond to the weakest observed AGN activity. 
The averaged fluxes measured in the six UVOT bands during these epochs were then used to estimate an upper limit on the host-galaxy contribution. 
Details of the five UVOT epochs with the lowest UVW2 fluxes, together with the host-galaxy fluxes estimated using different methods and the corresponding uncertainties are given in the Appendix (Table~\ref{tab:faintest5_and_host}).

Fig.~\ref{fig:variability_decomposition}c 
shows a comparison of the host galaxy spectra derived from decomposing thet data into a constant plus fixed shape variable component (cyan) versus that derived from the five faintest UVW2 spectra (red). Clearly these are most different in their UV predicted fluxes, with the faintest UVOT points having much more UV flux than predicted by the linear analysis. 
However, these two estimators cross around 3500\AA, so that the 
faintest UVOT points are slightly dimmer in B and V than predicted by the linear variability analysis. 

In the subsequent SED fitting, we 
choose to use the UVOT spectra from the 5 faintest UVW2 epochs as the most 
conservative estimation of the host-galaxy contamination, and use the absolute value of the 
difference between this and the host-galaxy estimated from the linear variability analysis as the systematic uncertainty of host estimation in each band, and propagate errors from this together with the statistical photometric uncertainties.

\section{Broadband spectral evolution}\label{Sec: sed}

We jointly fit the host galaxy subtracted UVOT and XRT spectra for each epoch to derive the broadband SED using the \texttt{AGNSED} model \citep{Kubota2018}. This is based physically on the Novikov-Thorne emissivity for a thin disc, but the spectral shape is assumed to be hot Comptonisation (for $R_{\rm isco}<R<R_{\rm h}$), or warm Comptonisation (for $R_{\rm h}<R<R_{\rm w}$), with blackbody only at $R>R_{\rm w}$. This radially stratified model gives enough flexibility to fit the spectrum, but with a physical constraint that these components are powered by the accretion flow, where the accretion rate is constant with radius. The hot Comptonisation is parameterised by spectral index $\Gamma_h$, and electron temperature $kT_{e,h}$ which is fixed at 100~keV as our fits do not include high energy data to constrain this. The warm Comptonisation is parameterised by spectral index $\Gamma_w$ and electron temperature $kT_{e,w}$. 

We fixed the Galactic hydrogen column density to $N_{\rm H}=7\times10^{19}\ {\rm cm^{-2}}$ \citep{HI4PI2016}. This is one of the lowest columns possible through our galaxy, giving one of the best views of the soft X-ray spectrum and its evolution during the event. There are also low column density ionised absorbers (and emission lines) seen in the XMM-Newton RGS data \citep{Parker2019}, but we ignore these here as the lower signal-to-noise Swift XRT is not sensitive to these. Similarly, the 
optical/UV reddening of $E(B-V)=0.008$ is so small that it makes no significant difference to even the UV filter fluxes so we neglect it in the fitting. 

The black-hole mass was fixed at $M_{\rm BH}=6\times10^{6}\ M_{\odot}$ \citep{Ochmann2024}, and the gravitational radius $R_g=GM_{\rm BH}/c^2$. We also fix the spin at zero, so the inner radius is $6R_g$, inclination at $\cos i=0.9$,
comoving distance at $D=21$~Mpc, the redshift at $z=0.005017$. 
These values were kept constant for all epochs, while the parameters describing the accretion flow were allowed to vary. 
The flow outer radius is assumed to be the self-gravity radius, $R_{sg}$ which changes with mass accretion rate as
$R_{sg}/R_g=2150 (M_{\rm BH}/10^9M_\odot)^{-2/9} \dot{m}^{4/9}\alpha^{2/9}\sim 4000\dot{m}^{4/9}$ for NGC1566 assuming viscosity $\alpha=0.1$ \citep{Laor1989}.

The {\sc agnsed} parameters for each broadband spectral fit are reported in Table~\ref{tab: sed_fitting}

\subsection{First peak}
\begin{figure*}[hb]
    \centering
    \includegraphics[width=\textwidth]{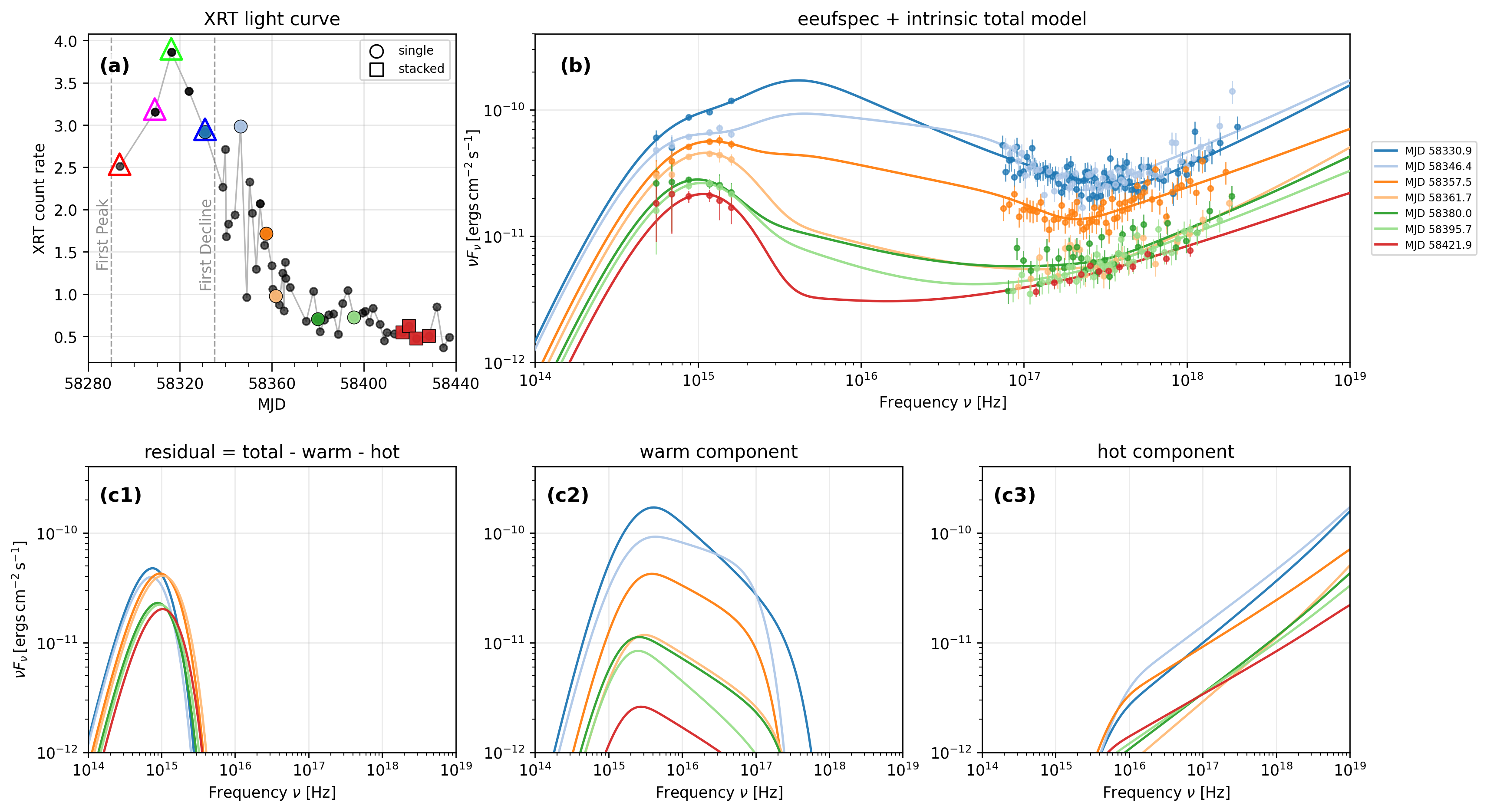}
    \caption{
    Evolution of the XRT light curve and broadband SED components of NGC~1566 between MJD 58280-58440. Panel (a) shows the Swift/XRT light curve, with the epochs used for SED fitting highlighted in colour. 
    Panel (b) presents the corresponding observed UVOT/XRT SEDs and best-fitting total \texttt{AGNSED} models. 
    Panels (c1)--(c3) decompose the observed model into the cold, warm, hot components. 
    }
    \label{fig:lightcurve_totalSED_components}
\end{figure*}

\begin{figure}
    \includegraphics[width=0.4\textwidth]{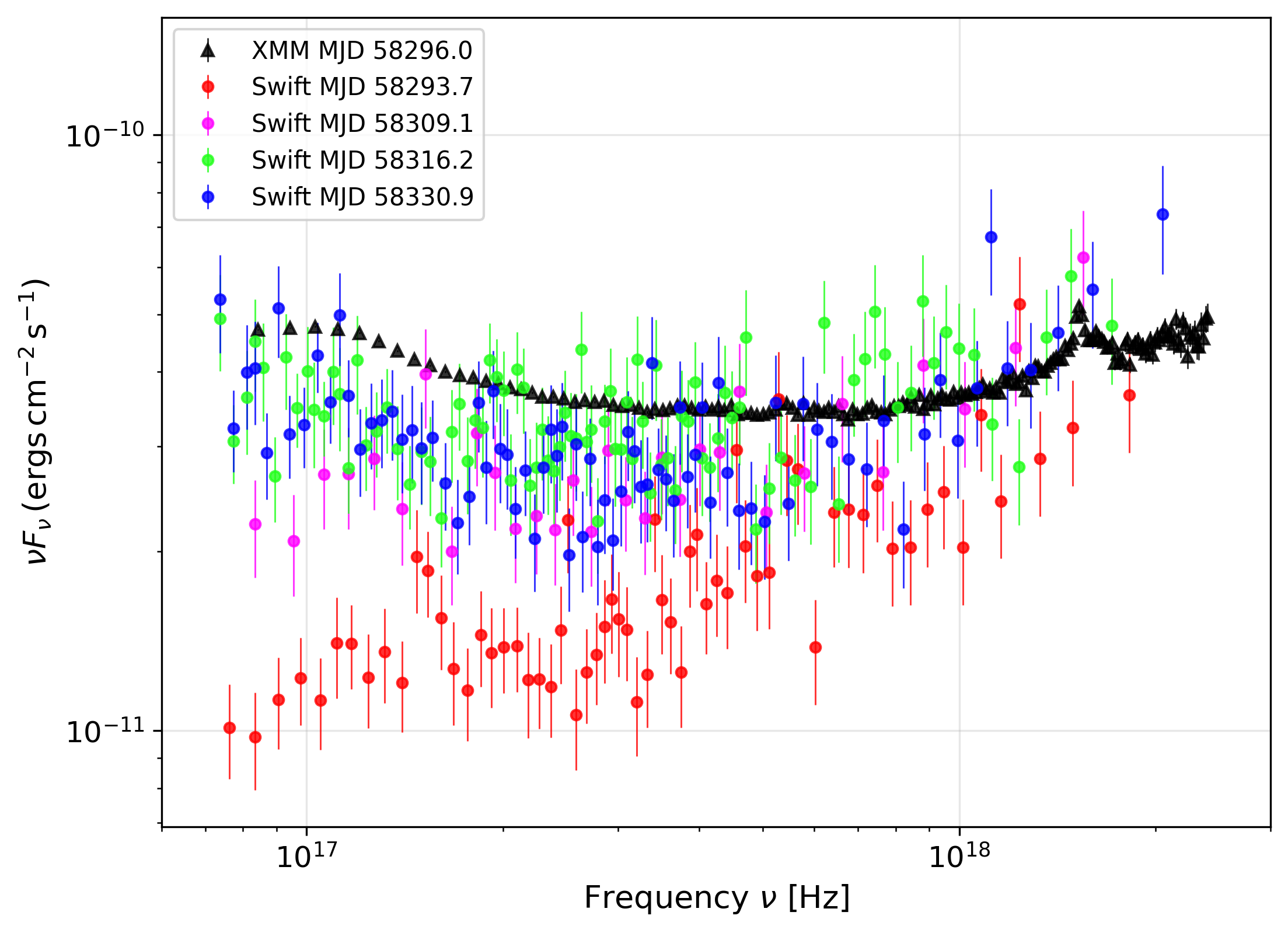}
    \caption{
    X-ray SED comparison of NGC1566 from XMM-Newton and Swift observations. 
    The black triangles represent the XMM-Newton spectrum at MJD58296.0, while the red, magenta, green and blue circles show the Swift spectra at MJD58293.8, MJD58309.1, MJD58316.2 and MJD58330.9 respectively. The swift epochs are marked in the light curve in Figure~\ref{fig:lightcurve_totalSED_components}a with triangular outlines in the same colours.
    The spectra are plotted in $\nu F_{\nu}$ as a function of frequency, allowing a direct comparison of the X-ray spectral shape and flux level between the XMM-Newton observation and the two nearby Swift epochs.
    }
    \label{fig:XMM_eeuf}
\end{figure}

Fig.~~\ref{fig:lightcurve_totalSED_components}a shows the \textit{Swift}/XRT lightcurve around the peak and the first decline. The point marked with the blue triangle is the first observation with full UVOT filter set. The four earlier time points have only UVW2 data, so these are not suitable for the broad band fits. Instead, we use these to explore the 
X-ray spectral evolution around the first peak. 

The spectra from the first three 
\textit{Swift}/XRT spectra (red, magenta and green open triangles in Fig.~\ref{fig:lightcurve_totalSED_components}a) are shown as red, magenta and green points in 
Fig.~\ref{fig:XMM_eeuf}. It is clear that these are similar at high energies, but the first (red) spectrum is a factor 3-4 lower at the softest energies. The change in spectral shape is made more evident by comparing to the XMM-Newton spectrum (black) taken between the first (red)  and second (magenta) datasets. The XMM-Newton data show a
clear rise at low energies associated with the soft X-ray excess, matching to the much lower signal to noise (shorter exposures as well as smaller effective area) \textit{Swift}/XRT spectra from the second (magenta) and third (green) observations. 
The first few \textit{Swift}/XRT spectra 
then directly show the appearance of the soft X-ray excess component. 

We also show for comparison the \textit{Swift}/XRT spectrum of the first dataset for which the full UVOT filter set is available (blue triangle in Fig.~\ref{fig:lightcurve_totalSED_components}a, and blue points in Fig.~\ref{fig:XMM_eeuf}). This is very similar to the other X-ray spectra around the peak, with curvature indicating the soft X-ray excess component is present.

\subsection{First decline}

Figure~\ref{fig:lightcurve_totalSED_components}b shows the broadband SED evolution
for selected epochs marked with filled coloured circles in  Figure~\ref{fig:lightcurve_totalSED_components}a. These span times from 
the first \textit{Swift}/XRT spectrum with full UVOT filter set which is close to the peak (blue, see also Fig.~\ref{fig:XMM_eeuf}), down through to its decline (red, with multiple points of the same colour indicating a stacked dataset).  

The first full \textit{Swift} broadband spectrum past the peak  (blue: Fig.~\ref{fig:lightcurve_totalSED_components}a,b) shows a strong UV/optical continuum together with a clear soft X-ray excess. In the model decomposition, the hard X-ray emission is dominated by the hot Comptonisation component, whereas the optical/UV and soft X-ray excess are fit mainly by the warm Comptonisation component, though with a small contribution inferred from the outer disc emission (see components in Fig.~\ref{fig:lightcurve_totalSED_components}c1-c3).

\begin{figure*}
    \centering
    \includegraphics[width=\textwidth]{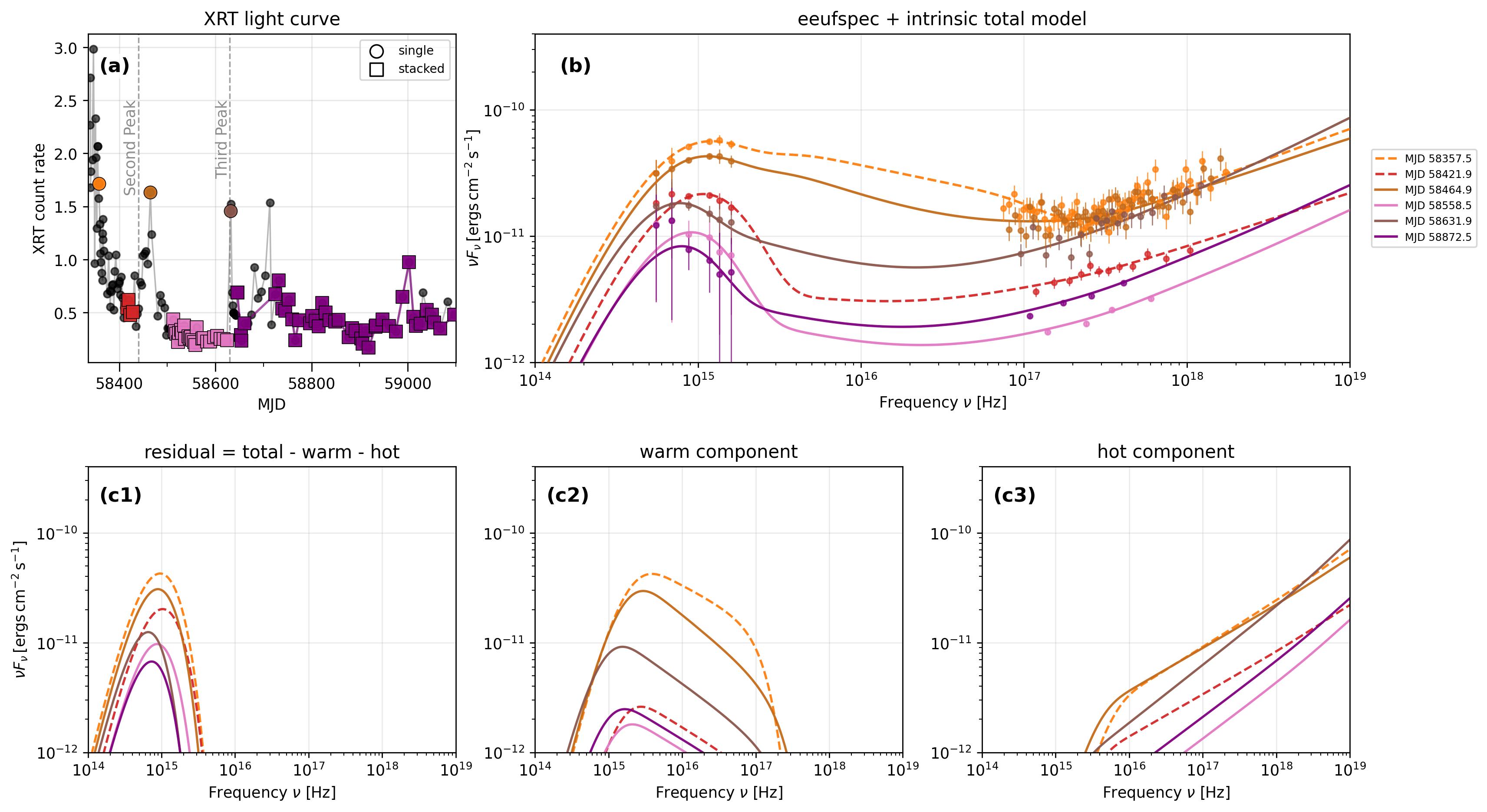}
    \caption{
    Evolution of the XRT light curve and broadband SED components of NGC~1566 between MJD 58335-59100. Panel (a) shows the Swift/XRT light curve, with the epochs used for SED fitting highlighted in colour. 
    Panel (b) presents the corresponding observed UVOT/XRT SEDs and best-fitting total \texttt{AGNSED} models. 
    Panels (c1)--(c3) decompose the observed model into the cold, warm, hot components. 
    }
    \label{fig:lightcurve_totalSED_components_low}
\end{figure*}

As the source fades from the peak, all
the wavebands decrease but with 
different amplitudes.
This is most clearly seen in the component decomposition: 
the hot Comptonisation component and the optical emission (modelled here as a residual outer disc) change much less than the warm Comptonisation component
(Fig.~\ref{fig:lightcurve_totalSED_components}c1-c3). 
The warm Comptonisation is clearly present in the brightest spectra (blue, light blue and orange: hereafter termed high state) but there is no clear soft X-ray excess or blue UV component from the yellow spectrum onwards
(hereafter low state: from MJD58362.1).
We keep the full {\sc agnsed} model i.e. allow the possibility of a warm Comptonisation region, but this is generally not significantly required by the low state datasets (see model details in Table~~\ref{tab: sed_fitting}). 
The modelled optical/UV emission changes from being dominated by the blue/UV warm Comptonisation in the high state, to being dominated by the redder outer disc in the low state, which  
explains why the linear variability analysis in Sec.~\ref{sec:host} fails, as the variable optical/UV component does not have a constant shape. 

While the UV and X-rays both clearly follow the decline, it is also clear that there is short timescale X-ray variability that is not linked with a similar change in optical/UV flux e.g. there is a fast flare on day MJD 58346.4 (light blue) where the X-rays are 50\% higher than the day before, and a factor 2 lower the day after. The X-ray spectrum is similar to that of the first SED (MJD 58330.9, blue), but the UVOT points are all lower. 
This shows that the intrinsic fast variability of the X-ray region does not propagate from and/or reprocess into the UV emission, but rather this is a separate region with intrinsic fast variabilty \citep{Hagen2025}.

\subsection{Second peak and beyond}

The remaining data consist of a series of smaller X-ray flares and long quiescent periods. The second peak (dark orange in Figure~~\ref{fig:lightcurve_totalSED_components_low}) is similar to the last high state spectrum on the first peak decline (dashed orange in Fig.~~\ref{fig:lightcurve_totalSED_components_low}, repeated from Fig.~~\ref{fig:lightcurve_totalSED_components} for completeness) but with less evidence for a soft X-ray excess. 

All other spectra around the second peak are low states at varying flux, so we show the full evolution of the SEDs across this period only in the Appendix.  

The third peak (brown) is at similar 
\textit{Swift}/XRT count rate as the second peak, but has even less evidence for a soft X-ray excess. This is a fast X-ray flare, and there is no strong UV change associated with this, suggesting again that the UV variability timescale is longer than the X-ray timescale. 

Figure~~\ref{fig:lightcurve_totalSED_components_low}a,b also shows the low state spectra from the second (magenta) and third decline (purple), compared to the first decline (red dashed line, repeated from Fig.~~\ref{fig:lightcurve_totalSED_components} for completeness). Although these are all low states, there are subtle differences. The second decline is a factor 2 lower in flux than the first decline, but has similar shape, while the third decline has similar flux to the second, but has a slightly different shape with weaker UV and stronger X-rays. The second and third low states are quite extended in time, so this is not just due to fast variability in the X-ray,
but instead 
implies some structural change in the optical/UV emission. 

\section{Origin of the optical/UV emission}

We now explore the origin of the optical/UV emission  in more detail across the entire outburst. 

\subsection{Diffuse continuum from the BLR}

In the high state, it is evident that the
data fit with {\sc agnsed} require both optical/UV emission from the outer blackbody disc, and an inner warm Comptonisation region. These are mainly produced by the intrinsic dissipation, but the model also includes additional heating of both these optically thick components from coronal X-ray illumination and subsequent thermalisation. 

However, recent progress from intensive broadband continuum monitoring campaigns has shown that reprocessing from optically thin material in the BLR and/or a wind inwards of the BLR is more significant than optically thick disc reprocessing, as seen by the long lag times and strong Balmer continuua (see e.g. \citet{Edelson2019}, \citet{Hagen2024}).

In NGC~1566 the
H$\beta$ FWHM of 2200~km/s \citep{Ochmann2024} gives an estimate for the 
BLR radius during the first bright state around $10^4R_g \sim 10^{16}$~cm, with a light-travel time of $\sim 1.7$ days, similar to that derived from the monochromatic continuum luminosity at 5100\AA~of $3.2\times 10^{42}$~ergs~s$^{-1}$
using \citet{Bentz2013}. We assume a cloud density of $10^{13}$~cm$^{-3}$ with column density of $10^{23}$~cm$^{-2}$ and covering fraction of $0.3$. 

We use {\sc cloudy} v25.00 \citep{Gunasekera2025} to calculate the diffuse continuum (DC) emission across the changing state event in NGC1566. We use the first high state with all UVOT filters (MJD58330.9, blue) and the low state on the first decline (MJD58421.9, red) to span the state transition. 

The characteristic BLR emission lines drop after the transition to the low state
\citep{Ochmann2024}, but this does not necessarily mean that the BLR itself disappears. The BLR emission will drop due to the drop in UV ionising flux, but the BLR structure  
should still be present, 
as this can only change on a 
dynamical timescale which is 
$\sim H/v_{FWHM}=150$~days, again assuming $H/R=0.3$ and $R=10^4R_g$ for the BLR. Hence we use the same BLR parameters for both the high and low state spectra to calculate the DC component. 

\begin{figure}
    \centering
    \includegraphics[width=\columnwidth]{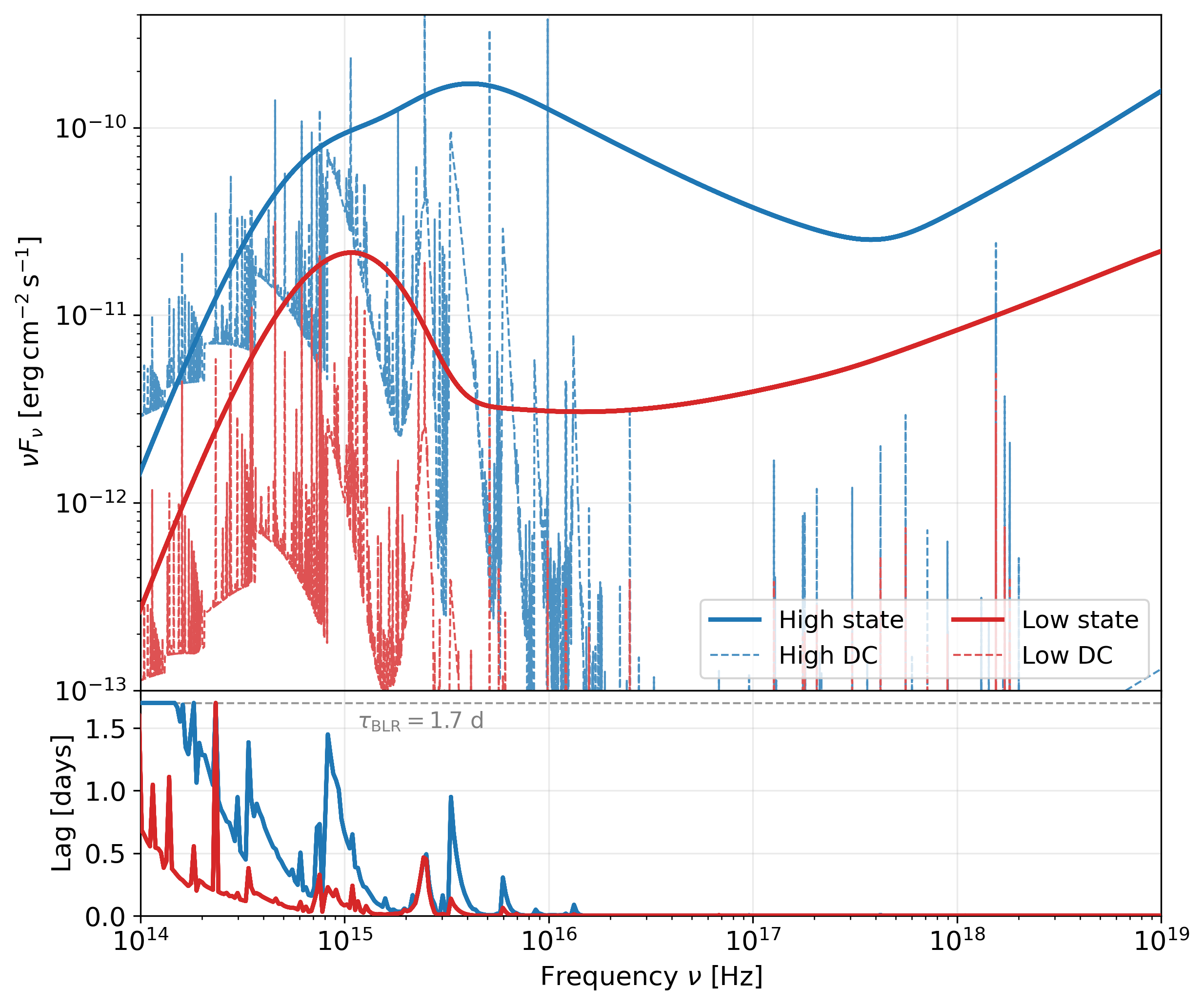}
    \caption{Comparison of the broadband SEDs and the corresponding {\sc{cloudy}} DC emission in the high and low states from the first peak and decline. The solid curves show the intrinsic \texttt{AGNSED} continuum, and the dashed lines show the {\sc{cloudy}} emission spectra. Blue and red denote the high and low states, respectively. The lower panel shows the wavelength-dependent continuum lag predicted for the two states, with the grey dashed horizontal line marking the estimated BLR light-travel time.
    }
    \label{fig:cloudy}
\end{figure}

Fig. \ref{fig:cloudy} shows the broadband SEDs (solid lines) and the corresponding {\sc cloudy} DC emission (dashed lines) for the high (blue) and low (red) state spectra. The bright high state spectrum has strong UV flux, so has a strong DC component which will follow the UV variable disc (typically longer timescale, and not well correlated with the fast variable X-ray coronal emission), as often seen in intensive continuum reverberation mapping campaigns of high state AGN \citep{Edelson2019}. The large size scale of the BLR compared to the disc also explains the result from these campaigns of longer lag times of the optical behind the UV than predicted by disc reprocessing \citep{Korista2001,Korista2019}.

The low state has much smaller DC contribution due to the drop in ionising UV/soft X-ray flux, 
but there is still some UV from the low energy extension of the X-ray coronal emission. This still gives some DC with its characteristic Balmer continuum emission, but this should now follow the faster X-ray variability, but again it is lagged and smoothed by the light travel time to the BLR radius.

We estimate the lag spectrum associated with the DC component from the BLR in high and low state by the fraction of DC at each wavelength, multiplied by the mean light travel time lag from the BLR of 1.7~days. 
This is shown in the lower panel of Fig.~\ref{fig:cloudy}.
This makes it clear that while the DC can still be present in the low state, its contribution to the optical/UV spectrum is much smaller due to the drop in intrinsic UV/soft X-ray ionising continuum. 
Thus the DC is not likely to make a large contribution to the low state spectra observed here.

\subsection{Disc reprocessing in the low states}

\begin{figure*}
    \centering
    \includegraphics[width=1.6\columnwidth]{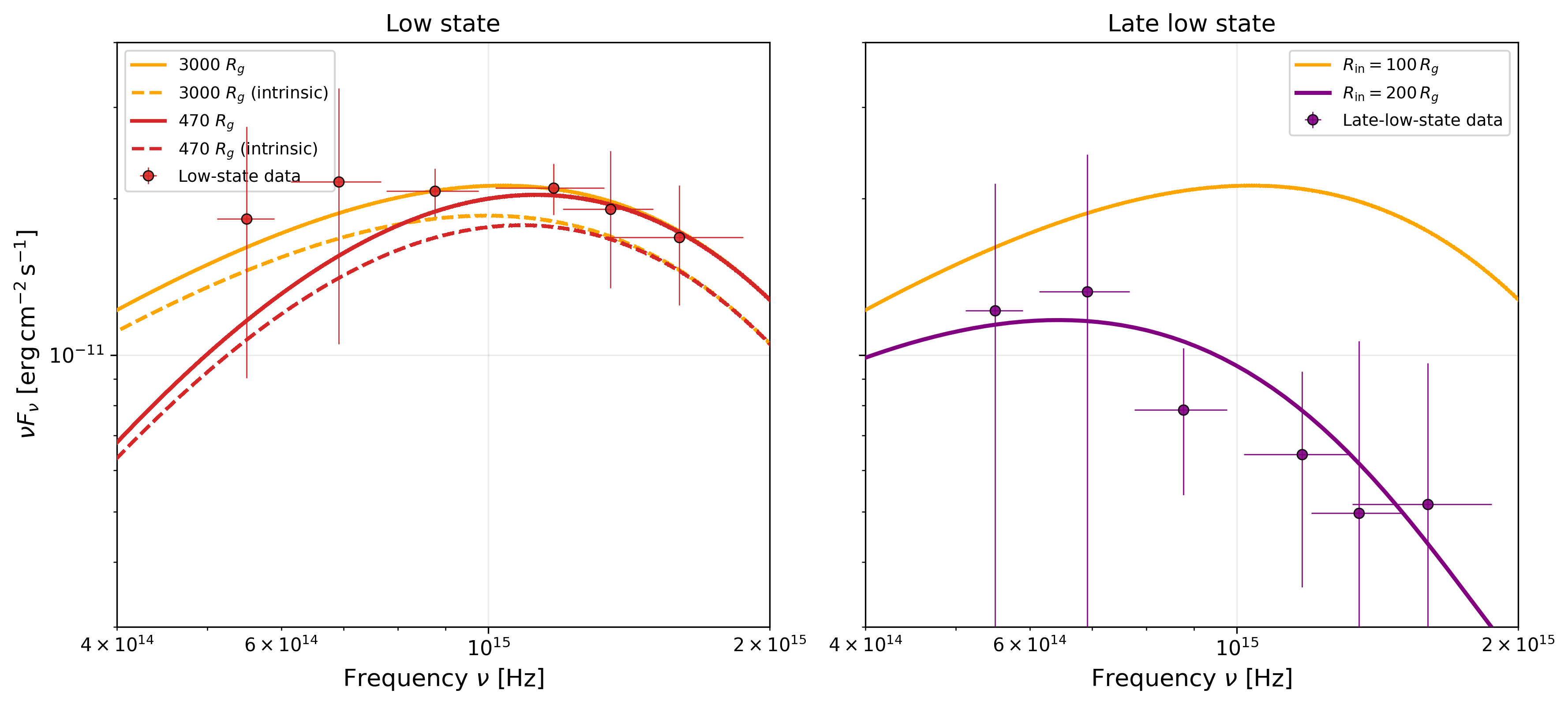}
    \caption{
    Comparison of the optical/UV SED predicted with different outer-disc radii.
    In left panel, the orange and red curves correspond to outer radii of $3000\,R_g$ and $470\,R_g$, respectively.
    In right panel, the orange and purple curves correspond to inner radii of $100\,R_g$ and $200\,R_g$, respectively.
    Solid lines show the reprocessed model spectra, while dashed lines show the corresponding intrinsic emission without irradiation.
    The black points show the observed UVOT fluxes with their $1\sigma$ uncertainties.
    }
    \label{fig:outerdisc_comparison}
\end{figure*}

We show the UVOT data from the lowest spectrum on the first decline (red points), together with the best fit 
{\sc AGNSED} model from Section 4.2 (red solid line) in Fig.~\ref{fig:outerdisc_comparison}. It is clear that this underestimates the V and B band fluxes. Closer inspection of all the 
low state data  
(dark green onwards in the decline from the first peak
Figs.~\ref{fig:lightcurve_totalSED_components}
as well as the low states shown in Figs.~\ref{fig:lightcurve_totalSED_components_low}) show that this is a common problem across all low state spectra. The previous section shows that this is not likely to be from a strong DC component as the drop in UV flux means that the DC component should contribute less to the optical emission. 

Intead, we explore X-ray irradiation of the disc. 
The  {\sc agnsed} modeling has the assumption that the outer disc radius is set by self gravity \citep{Kubota2018}, which is $\sim 470R_g$ for these data (Section~~\ref{Sec: sed}). However, this is calculated assuming a Shakura-Sunyaev (radiation pressure dominated) disc, which is not consistent with the large ($\sim 100R_g$) hot Comptonising corona region derived in the model fit, nor with the fact that the disc emission varies so rapidly. Instead we assume that the optically thick disc can extend outwards to $10^4R_g$, approaching broad line region (BLR) scales. The orange solid line in Fig.~\ref{fig:outerdisc_comparison} shows how this increases the model V and B band fluxes, in much better agreement with the data. This includes the contribution of X-ray reprocessed flux 
from the hot Comptonisation region assuming a scale height of $10R_g$ irradiating a flat, optically thick disc \citep{Kubota2018}. 
The dashed lines in Fig.~\ref{fig:outerdisc_comparison}a show the intrinsic disc emission without reprocessing. 
This disc reprocessed component should follow the X-ray variability with a lag which is characteristic of the $\sim 100R_g$ inner disc radius \citep{Hagen2023}.

\begin{figure}
    \centering
    \includegraphics[width=0.6\columnwidth]{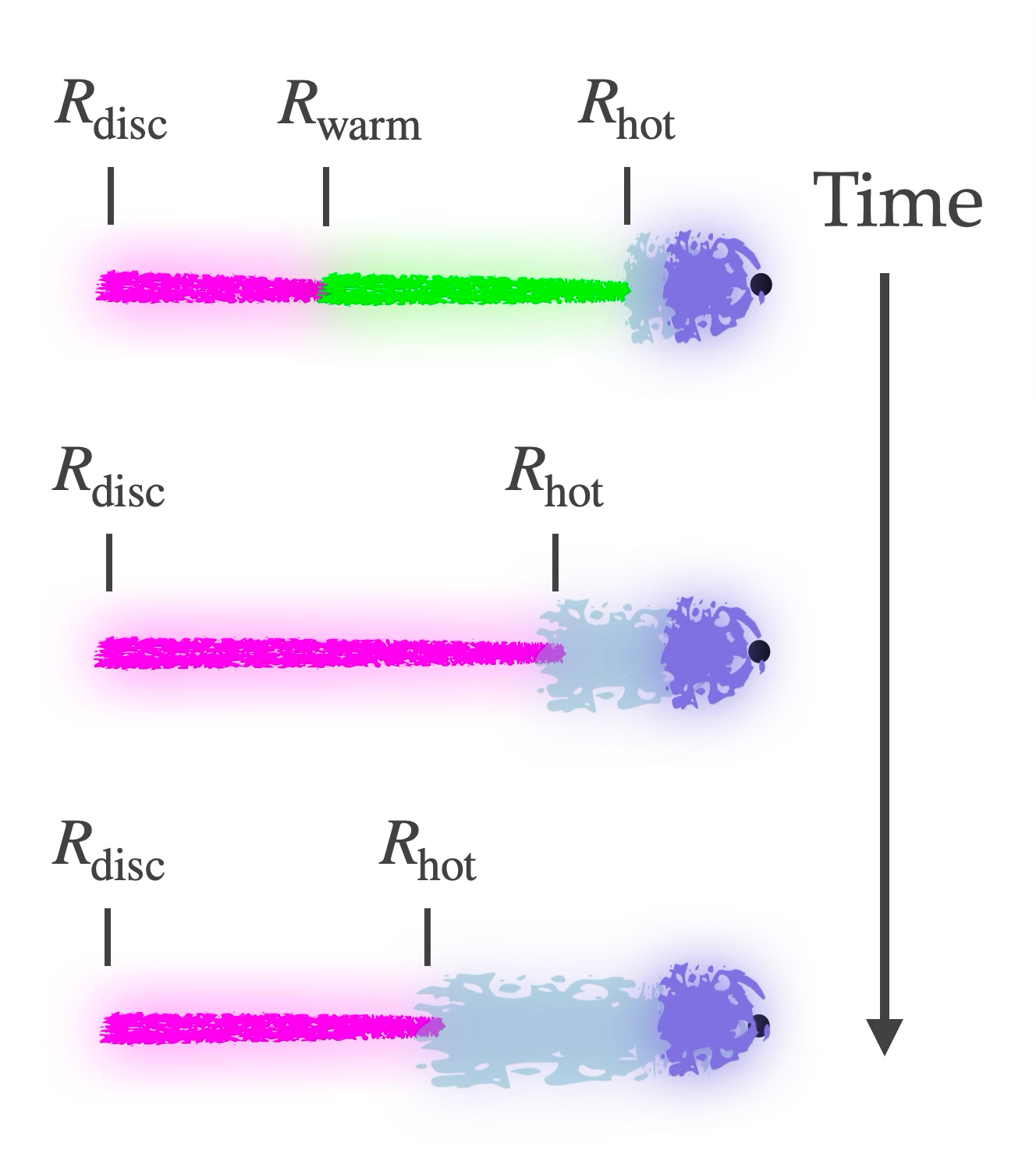}
    \caption{Schematic illustration of the evolution of the accretion flow from the high state to the low and late-low state, showing the hot corona (blue/light blue), warm corona (green) and cold accretion disc (magenta). 
    }
    \label{fig:outerdisc_model}
\end{figure}

The low states on the third decline have a slightly different optical/UV shape to the low states on the first and second declines (see 
Fig.\ref{fig:lightcurve_totalSED_components_low}). We show these data in 
Fig.~\ref{fig:outerdisc_comparison}b
and it is clear that these spectra are dimmer in the UV, but similar in the optical. The UV is emitted at the inner edge of the truncated disc in the {\sc agnsed} model, so this  
can be matched by the truncated disc receeding progressively outwards. We fit this and find typical radii of $R_h\sim 100R_g$ for the first decline (red), expanding to $\sim 280R_g$ for the third decline (magenta). 
There is evidence for a similarly receeding truncation radius in low state  HST spectra from NGC3147, an AGN at $L/L_{\rm Edd}\sim 10^{-4}$ \citep{Bianchi2022}. 

\section{Long term behaviour of  NGC1566}

NGC1566 is a repeating CS-AGN, with historic spectroscopic BLR detections in 1963, 1982 and 1987 and inferred BLR from similarly high optical/IR spectra in 1961, 1976 and possibly 2006 (the abrupt change in Catalina  Real-Time
Transient Survey (CRTS, \citealt{Drake2009}) optical flux looks suspicious). These are detailed in  the 
Appendix (Fig.~\ref{fig:ngc1566_longterm_history} and references therein). We also show a  zoom-in of the 2016-2026 optical (V and g band) 
ASAS-SN (\citealt{Hart2023})
lightcurve 
(Fig.~\ref{fig:asassn_lc}). This reveals that the large outburst in 2018 appears quite symmetric, with clear short bursts of actvity on a 100 day baseline before the Swift monitoring started. 

The SED fits in Section 4 show that the second peak is the lowest luminosity spectrum which has clear evidence for the warm component, so we take this marking the transition. This corresponds to an ASAS-SN g-band magnitude (which includes a substantial contribution from the host galaxy) of $12.2$~mag. We encourage using this criteria as a trigger for broadband and spectroscopic observations. 

The ASAS-SN lightcurve also 
reveals another small peak in 2023, seen also in the ATLAS forced photometry difference images, but this is below 12.2~mag, so probably did not trigger a transition. 

The sharp events seen in NGC1566 are quite unlike the more typical slower stochastic rise/fall seen in other repeat CS-AGN (see e.g. the review by \citealt{komossa2026}). The only other comparable object is IC3599, though here the events are even larger amplitude, probably reaching Eddington
\citep{Grupe2026}. Perturbation of the accretion flow by a binary black hole companion \citep{Dodd2025} 
is clearly an attractive scenario
for such events, but the data do not support a clear periodicity. Nontheless, this could induce more complex dynamics (spiral arms in the accretion flow \citealt{Dodd2025} and/or an ellipsoidal disc \citealt{Ochmann2024} and/or disc tearing events \citealt{Raj2021}) which might make this plausible. 

\section{Summary}\label{Sec: summary}

We re-analyze the 
optical/UV and X-ray monitoring of the 
2018 outburst of 
NGC~1566, a low mass repeating CS-AGN, 
with $M_{\rm BH}\sim6\times 
10^{6}M_{\odot}$. This shows the well known changing state behaviour, with the strong UV/soft X-ray component dissapearing around $L\sim 0.01L_{\rm Edd}$. This dramatic drop in ionising photons causes the disappearance of the BLR emission, though the BLR structure itself can only vary on a longer (dynamical) timescale. 

With careful host galaxy subtraction, we show the evolution of the optical/UV spectrum across the transition and into the low state. There is clear evidence for a red optical component in both high and low states, probably from reprocesssing
on BLR scales as well as from direct and reprocessed emission from the outer disc.
The BLR has a larger scale height, but is likely optically thin (column of $10^{23}$~cm$^{-2}$) so effectively reprocesses only UV/soft X-ray flux. Conversely, the disc is optically thick so is a more effective reprocessor of the hard X-ray flux. The strong ionising UV flux in the bright states means that the BLR reprocessing dominates, so intensive continuum monitoring campaigns should be dominated by the BLR scale lagged diffuse continuum, whereas the drop in this flux at the CS transition means that the UV is dominated by the X-ray coronal power law, so the BLR diffuse continuum now follows the X-ray variability but is still lagged on BLR scales. The deep decline states may instead be dominated by X-ray reprocessing on the truncated disc, with inner radius expanding from $100-250R_g$. 

NGC1566 has shown repeated CS events associated with strong distinct flares in its lightcurve. The source is continuually monitored (by Atlas, ASAS-SN and now LSST) and we strongly encourage use of these to trigger observations of new events. One very attractive possibility is that there is a binary companion, perhaps an intermediate mass black hole, perturbing the flow \citep{Dodd2025}.

CS-AGN give unique insights into the changing nature of the accretion flow, and we will explore the impact of these on our understanding of bright quasar spectra
in a subsequent paper (Wang et al 2026, in prep). In 
particular we will use them to assess the scatter in the $L_{UV}-L_X$ relation, and hence evaluate whether these can be used as 
standardizable candles to give distance independent from redshift, enabling them to be used for cosmology \citep{Lusso2016,Lusso2020}.

\section*{Acknowledgments}
CD acknowledges support from STFC through the
grant ST/T000244/1. RM acknowledges financial support from the INAF Scientific Directorate. XBW thanks the support of the National Key R\&D Program of China (2025YFA1614100, 2025YFA1614101) and
the National Science Foundation of China (12133001, 12633003).

This work made use of data supplied by the UK Swift Science Data Centre at the University of Leicester
and the High Energy Astrophysics Science Archive Research Center (HEASARC), which is a service of the Astrophysics Science Division at NASA/GSFC. The Catalina Sky Survey is funded by the National Aeronautics and Space Administration under Grant No. NNG05GF22G issued through the Science Mission Directorate Near-Earth Objects Observations Program. This research has made use of data from the All-Sky Automated Survey for SuperNovae (ASAS-SN). ASAS-SN is supported by the Gordon and Betty Moore Foundation through grant GBMF5490 to the Ohio State University, and by the National Science Foundation under Grant No. AST-1515921. We thank the Las Cumbres Observatory and its staff for its continuing support of the ASAS-SN project.

\section*{Data Availability}

The optical/UV and X-ray data from Swift and XMM-Newton data used in this work are publicly available from their respective archives. The optical lightcurves showing the long term behaviour are publically available from ASAS-SN and CSS.

\bibliographystyle{mnras}
\bibliography{sample631}{}

@ARTICLE{Middei2022,
       author = {{Middei}, R. and {Marinucci}, A. and {Braito}, V. and {Bianchi}, S. and {De Marco}, B. and {Luminari}, A. and {Matt}, G. and {Nardini}, E. and {Perri}, M. and {Reeves}, J.~N. and {Vagnetti}, F.},
        title = "{The lively accretion disc in NGC 2992 - II. The 2019/2021 X-ray monitoring campaigns}",
      journal = {\mnras},
         year = 2022,
        month = aug,
       volume = {514},
       number = {2},
        pages = {2974-2993},
          doi = {10.1093/mnras/stac1381},
archivePrefix = {arXiv},
       eprint = {2205.07904},
 primaryClass = {astro-ph.HE},
       adsurl = {https://ui.adsabs.harvard.edu/abs/2022MNRAS.514.2974M}
}

@ARTICLE{Evans2009,
       author = {{Evans}, P.~A. and {Beardmore}, A.~P. and {Page}, K.~L. and {Osborne}, J.~P. and {O'Brien}, P.~T. and {Willingale}, R. and {Starling}, R.~L.~C. and {Burrows}, D.~N. and {Godet}, O. and {Vetere}, L. and {Racusin}, J. and {Goad}, M.~R. and {Wiersema}, K. and {Angelini}, L. and {Capalbi}, M. and {Chincarini}, G. and {Gehrels}, N. and {Kennea}, J.~A. and {Margutti}, R. and {Morris}, D.~C. and {Mountford}, C.~J. and {Pagani}, C. and {Perri}, M. and {Romano}, P. and {Tanvir}, N.},
        title = "{Methods and results of an automatic analysis of a complete sample of Swift-XRT observations of GRBs}",
      journal = {\mnras},
         year = 2009,
        month = aug,
       volume = {397},
       number = {3},
        pages = {1177-1201},
          doi = {10.1111/j.1365-2966.2009.14913.x},
archivePrefix = {arXiv},
       eprint = {0812.3662},
 primaryClass = {astro-ph},
       adsurl = {https://ui.adsabs.harvard.edu/abs/2009MNRAS.397.1177E}
}

@ARTICLE{Kynoch2026,
       author = {{Kynoch}, D. and {McHardy}, I.~M. and {Cackett}, E.~M. and {Gelbord}, J. and {Hern{\'a}ndez Santisteban}, J.~V. and {Horne}, K. and {Miller}, J.~A. and {Netzer}, H. and {Done}, C. and {Edelson}, R. and {Fausnaugh}, M.~M. and {Goad}, M.~R. and {Peterson}, B.~M. and {Vincentelli}, F.~M.},
        title = "{Intensive X-Ray/UVOIR continuum reverberation mapping of the Seyfert AGN MCG +08─11─11}",
      journal = {\mnras},
         year = 2026,
        month = mar,
       volume = {546},
       number = {3},
          eid = {stag025},
        pages = {stag025},
          doi = {10.1093/mnras/stag025},
archivePrefix = {arXiv},
       eprint = {2511.05342},
 primaryClass = {astro-ph.GA},
       adsurl = {https://ui.adsabs.harvard.edu/abs/2026MNRAS.546ag025K}
}

@ARTICLE{Peng2002,
       author = {{Peng}, Chien Y. and {Ho}, Luis C. and {Impey}, Chris D. and {Rix}, Hans-Walter},
        title = "{Detailed Structural Decomposition of Galaxy Images}",
      journal = {\aj},
         year = 2002,
        month = jul,
       volume = {124},
       number = {1},
        pages = {266-293},
          doi = {10.1086/340952},
archivePrefix = {arXiv},
       eprint = {astro-ph/0204182},
 primaryClass = {astro-ph},
       adsurl = {https://ui.adsabs.harvard.edu/abs/2002AJ....124..266P}
}

@article{Done2012,
    author = {Done, Chris and Davis, S. W. and Jin, C. and Blaes, O. and Ward, M.},
    title = {Intrinsic disc emission and the soft X-ray excess in active galactic nuclei},
    journal = {Monthly Notices of the Royal Astronomical Society},
    volume = {420},
    number = {3},
    pages = {1848-1860},
    year = {2012},
    month = {03},
    issn = {0035-8711},
    doi = {10.1111/j.1365-2966.2011.19779.x},
    url = {https://doi.org/10.1111/j.1365-2966.2011.19779.x},
    eprint = {https://academic.oup.com/mnras/article-pdf/420/3/1848/3001571/mnras0420-1848.pdf},
}

@ARTICLE{Kubota2018,
       author = {{Kubota}, Aya and {Done}, Chris},
        title = "{A physical model of the broad-band continuum of AGN and its implications for the UV/X relation and optical variability}",
      journal = {\mnras},
         year = 2018,
        month = oct,
       volume = {480},
       number = {1},
        pages = {1247-1262},
          doi = {10.1093/mnras/sty1890},
archivePrefix = {arXiv},
       eprint = {1804.00171},
 primaryClass = {astro-ph.HE},
       adsurl = {https://ui.adsabs.harvard.edu/abs/2018MNRAS.480.1247K}
}

@ARTICLE{LaMassa2015,
       author = {{LaMassa}, Stephanie M. and {Cales}, Sabrina and {Moran}, Edward C. and {Myers}, Adam D. and {Richards}, Gordon T. and {Eracleous}, Michael and {Heckman}, Timothy M. and {Gallo}, Luigi and {Urry}, C. Megan},
        title = "{The Discovery of the First {\textquotedblleft}Changing Look{\textquotedblright} Quasar: New Insights Into the Physics and Phenomenology of Active Galactic Nucleus}",
      journal = {\apj},
         year = 2015,
        month = feb,
       volume = {800},
       number = {2},
          eid = {144},
        pages = {144},
          doi = {10.1088/0004-637X/800/2/144},
archivePrefix = {arXiv},
       eprint = {1412.2136},
 primaryClass = {astro-ph.GA},
       adsurl = {https://ui.adsabs.harvard.edu/abs/2015ApJ...800..144L}
}

@ARTICLE{MacLeod2016,
       author = {{MacLeod}, Chelsea L. and {Ross}, Nicholas P. and {Lawrence}, Andy and {Goad}, Mike and {Horne}, Keith and {Burgett}, William and {Chambers}, Ken C. and {Flewelling}, Heather and {Hodapp}, Klaus and {Kaiser}, Nick and {Magnier}, Eugene and {Wainscoat}, Richard and {Waters}, Christopher},
        title = "{A systematic search for changing-look quasars in SDSS}",
      journal = {\mnras},
         year = 2016,
        month = mar,
       volume = {457},
       number = {1},
        pages = {389-404},
          doi = {10.1093/mnras/stv2997},
archivePrefix = {arXiv},
       eprint = {1509.08393},
 primaryClass = {astro-ph.GA},
       adsurl = {https://ui.adsabs.harvard.edu/abs/2016MNRAS.457..389M}
}

@ARTICLE{Ruan2019,
       author = {{Ruan}, John J. and {Anderson}, Scott F. and {Eracleous}, Michael and {Green}, Paul J. and {Haggard}, Daryl and {MacLeod}, Chelsea L. and {Runnoe}, Jessie C. and {Sobolewska}, Malgosia A.},
        title = "{The Analogous Structure of Accretion Flows in Supermassive and Stellar Mass Black Holes: New Insights from Faded Changing-look Quasars}",
      journal = {\apj},
         year = 2019,
        month = sep,
       volume = {883},
       number = {1},
          eid = {76},
        pages = {76},
          doi = {10.3847/1538-4357/ab3c1a},
archivePrefix = {arXiv},
       eprint = {1903.02553},
 primaryClass = {astro-ph.HE},
       adsurl = {https://ui.adsabs.harvard.edu/abs/2019ApJ...883...76R}
}

@ARTICLE{Sheng2019,
       author = {{Sheng}, Zhenfeng and {Wang}, Tinggui and {Jiang}, Ning and {Ding}, Jiani and {Cai}, Zheng and {Guo}, Hengxiao and {Sun}, Luming and {Dou}, Liming and {Yang}, Chenwei},
        title = "{Initial Results from a Systematic Search for Changing-look Active Galactic Nuclei Selected via Mid-infrared Variability}",
      journal = {\apj},
         year = 2020,
        month = jan,
       volume = {889},
       number = {1},
          eid = {46},
        pages = {46},
          doi = {10.3847/1538-4357/ab5af9},
archivePrefix = {arXiv},
       eprint = {1905.02904},
 primaryClass = {astro-ph.GA},
       adsurl = {https://ui.adsabs.harvard.edu/abs/2020ApJ...889...46S}
}

@ARTICLE{MacLeod2019,
       author = {{MacLeod}, Chelsea L. and {Green}, Paul J. and {Anderson}, Scott F. and {Bruce}, Alastair and {Eracleous}, Michael and {Graham}, Matthew and {Homan}, David and {Lawrence}, Andy and {LeBleu}, Amy and {Ross}, Nicholas P. and {Ruan}, John J. and {Runnoe}, Jessie and {Stern}, Daniel and {Burgett}, William and {Chambers}, Kenneth C. and {Kaiser}, Nick and {Magnier}, Eugene and {Metcalfe}, Nigel},
        title = "{Changing-look Quasar Candidates: First Results from Follow-up Spectroscopy of Highly Optically Variable Quasars}",
      journal = {\apj},
         year = 2019,
        month = mar,
       volume = {874},
       number = {1},
          eid = {8},
        pages = {8},
          doi = {10.3847/1538-4357/ab05e2},
archivePrefix = {arXiv},
       eprint = {1810.00087},
 primaryClass = {astro-ph.GA},
       adsurl = {https://ui.adsabs.harvard.edu/abs/2019ApJ...874....8M}
}

@ARTICLE{Yang2018,
       author = {{Yang}, Qian and {Wu}, Xue-Bing and {Fan}, Xiaohui and {Jiang}, Linhua and {McGreer}, Ian and {Shangguan}, Jinyi and {Yao}, Su and {Wang}, Bingquan and {Joshi}, Ravi and {Green}, Richard and {Wang}, Feige and {Feng}, Xiaotong and {Fu}, Yuming and {Yang}, Jinyi and {Liu}, Yuanqi},
        title = "{Discovery of 21 New Changing-look AGNs in the Northern Sky}",
      journal = {\apj},
         year = 2018,
        month = aug,
       volume = {862},
       number = {2},
          eid = {109},
        pages = {109},
          doi = {10.3847/1538-4357/aaca3a},
archivePrefix = {arXiv},
       eprint = {1711.08122},
 primaryClass = {astro-ph.GA},
       adsurl = {https://ui.adsabs.harvard.edu/abs/2018ApJ...862..109Y}
}

@ARTICLE{Sheng2017,
       author = {{Sheng}, Zhenfeng and {Wang}, Tinggui and {Jiang}, Ning and {Yang}, Chenwei and {Yan}, Lin and {Dou}, Liming and {Peng}, Bo},
        title = "{Mid-infrared Variability of Changing-look AGNs}",
      journal = {\apjl},
         year = 2017,
        month = sep,
       volume = {846},
       number = {1},
          eid = {L7},
        pages = {L7},
          doi = {10.3847/2041-8213/aa85de},
archivePrefix = {arXiv},
       eprint = {1707.02686},
 primaryClass = {astro-ph.GA},
       adsurl = {https://ui.adsabs.harvard.edu/abs/2017ApJ...846L...7S}
}

@ARTICLE{Narayan1995,
       author = {{Narayan}, Ramesh and {Yi}, Insu},
        title = "{Advection-dominated Accretion: Underfed Black Holes and Neutron Stars}",
      journal = {\apj},
         year = 1995,
        month = oct,
       volume = {452},
        pages = {710},
          doi = {10.1086/176343},
archivePrefix = {arXiv},
       eprint = {astro-ph/9411059},
 primaryClass = {astro-ph},
       adsurl = {https://ui.adsabs.harvard.edu/abs/1995ApJ...452..710N}
}

@ARTICLE{Noda2018,
       author = {{Noda}, Hirofumi and {Done}, Chris},
        title = "{Explaining changing-look AGN with state transition triggered by rapid mass accretion rate drop}",
      journal = {\mnras},
         year = 2018,
        month = nov,
       volume = {480},
       number = {3},
        pages = {3898-3906},
          doi = {10.1093/mnras/sty2032},
archivePrefix = {arXiv},
       eprint = {1805.07873},
 primaryClass = {astro-ph.GA},
       adsurl = {https://ui.adsabs.harvard.edu/abs/2018MNRAS.480.3898N}
}

@ARTICLE{Middei2026,
       author = {{Middei}, R. and {Nardini}, E. and {Done}, C. and {Lusso}, E. and {Vagnetti}, F. and {Risaliti}, G. and {Piconcelli}, E. and {Bianchi}, S. and {Matzeu}, G. and {Trindade Falc{\~a}o}, A. and {Kr{\'o}l}, D. {\L}. and {Perri}, M. and {Maselli}, A. and {Horne}, K. and {Hern{\'a}ndez Santisteban}, J.~V.},
        title = "{Directly tracking the re-brightening of a supermassive black hole accretion disk}",
      journal = {arXiv e-prints},
         year = 2026,
        month = may,
          eid = {arXiv:2605.18958},
        pages = {arXiv:2605.18958},
          doi = {10.48550/arXiv.2605.18958},
archivePrefix = {arXiv},
       eprint = {2605.18958},
 primaryClass = {astro-ph.HE},
       adsurl = {https://ui.adsabs.harvard.edu/abs/2026arXiv260518958M}
}

@ARTICLE{Kang2025,
       author = {{Kang}, Jia-Lai and {Done}, Chris and {Hagen}, Scott and {Temple}, Matthew J. and {Silverman}, John D. and {Li}, Junyao and {Liu}, Teng},
        title = "{Systematic collapse of the accretion disc in AGN confirmed by UV photometry and broad-line spectra}",
      journal = {\mnras},
         year = 2025,
        month = mar,
       volume = {538},
       number = {1},
        pages = {121-131},
          doi = {10.1093/mnras/staf145},
archivePrefix = {arXiv},
       eprint = {2410.06730},
 primaryClass = {astro-ph.HE},
       adsurl = {https://ui.adsabs.harvard.edu/abs/2025MNRAS.538..121K}
}

@ARTICLE{Hagen2024,
       author = {{Hagen}, Scott and {Done}, Chris and {Edelson}, Rick},
        title = "{What drives the variability in AGN? Explaining the UV-Xray disconnect through propagating fluctuations}",
      journal = {\mnras},
         year = 2024,
        month = jun,
       volume = {530},
       number = {4},
        pages = {4850-4867},
          doi = {10.1093/mnras/stae1177},
archivePrefix = {arXiv},
       eprint = {2401.03452},
 primaryClass = {astro-ph.HE},
       adsurl = {https://ui.adsabs.harvard.edu/abs/2024MNRAS.530.4850H}
}

@ARTICLE{Palit2025,
       author = {{Palit}, Biswaraj and {{\'S}niegowska}, Marzena and {Markowitz}, Alex and {R{\'o}{\.z}a{\'n}ska}, Agata and {Farah}, Joseph and {Howell}, D. Andrew},
        title = "{Markarian 590: the AGN awakens}",
      journal = {\mnras},
         year = 2025,
        month = jun,
       volume = {540},
       number = {1},
        pages = {L14-L20},
          doi = {10.1093/mnrasl/slaf027},
archivePrefix = {arXiv},
       eprint = {2501.07225},
 primaryClass = {astro-ph.HE},
       adsurl = {https://ui.adsabs.harvard.edu/abs/2025MNRAS.540L..14P}
}

@ARTICLE{Lusso2016,
       author = {{Lusso}, E. and {Risaliti}, G.},
        title = "{The Tight Relation between X-Ray and Ultraviolet Luminosity of Quasars}",
      journal = {\apj},
         year = 2016,
        month = mar,
       volume = {819},
       number = {2},
          eid = {154},
        pages = {154},
          doi = {10.3847/0004-637X/819/2/154},
archivePrefix = {arXiv},
       eprint = {1602.01090},
 primaryClass = {astro-ph.GA},
       adsurl = {https://ui.adsabs.harvard.edu/abs/2016ApJ...819..154L}
}

@ARTICLE{Risaliti2002,
       author = {{Risaliti}, G. and {Elvis}, M. and {Nicastro}, F.},
        title = "{Ubiquitous Variability of X-Ray-absorbing Column Densities in Seyfert 2 Galaxies}",
      journal = {\apj},
         year = 2002,
        month = may,
       volume = {571},
       number = {1},
        pages = {234-246},
          doi = {10.1086/324146},
archivePrefix = {arXiv},
       eprint = {astro-ph/0107510},
 primaryClass = {astro-ph},
       adsurl = {https://ui.adsabs.harvard.edu/abs/2002ApJ...571..234R}
}

@ARTICLE{Petrucci2018,
       author = {{Petrucci}, P.-O. and {Ursini}, F. and {De Rosa}, A. and {Bianchi}, S. and {Cappi}, M. and {Matt}, G. and {Dadina}, M. and {Malzac}, J.},
        title = "{Testing warm Comptonization models for the origin of the soft X-ray excess in AGNs}",
      journal = {\aap},
         year = 2018,
        month = mar,
       volume = {611},
          eid = {A59},
        pages = {A59},
          doi = {10.1051/0004-6361/201731580},
archivePrefix = {arXiv},
       eprint = {1710.04940},
 primaryClass = {astro-ph.HE},
       adsurl = {https://ui.adsabs.harvard.edu/abs/2018A&A...611A..59P}
}

@ARTICLE{Crummy2006,
       author = {{Crummy}, J. and {Fabian}, A.~C. and {Gallo}, L. and {Ross}, R.~R.},
        title = "{An explanation for the soft X-ray excess in active galactic nuclei}",
      journal = {\mnras},
         year = 2006,
        month = feb,
       volume = {365},
       number = {4},
        pages = {1067-1081},
          doi = {10.1111/j.1365-2966.2005.09844.x},
archivePrefix = {arXiv},
       eprint = {astro-ph/0511457},
 primaryClass = {astro-ph},
       adsurl = {https://ui.adsabs.harvard.edu/abs/2006MNRAS.365.1067C}
}

@ARTICLE{Lusso2020,
       author = {{Lusso}, E. and {Risaliti}, G. and {Nardini}, E. and {Bargiacchi}, G. and {Benetti}, M. and {Bisogni}, S. and {Capozziello}, S. and {Civano}, F. and {Eggleston}, L. and {Elvis}, M. and {Fabbiano}, G. and {Gilli}, R. and {Marconi}, A. and {Paolillo}, M. and {Piedipalumbo}, E. and {Salvestrini}, F. and {Signorini}, M. and {Vignali}, C.},
        title = "{Quasars as standard candles. III. Validation of a new sample for cosmological studies}",
      journal = {\aap},
         year = 2020,
        month = oct,
       volume = {642},
          eid = {A150},
        pages = {A150},
          doi = {10.1051/0004-6361/202038899},
archivePrefix = {arXiv},
       eprint = {2008.08586},
 primaryClass = {astro-ph.GA},
       adsurl = {https://ui.adsabs.harvard.edu/abs/2020A&A...642A.150L}
}

@ARTICLE{Alloin1985,
       author = {{Alloin}, D. and {Pelat}, D. and {Phillips}, M. and {Whittle}, M.},
        title = "{Recent spectral variations in the active nucleus of NGC 1566.}",
      journal = {\apj},
         year = 1985,
        month = jan,
       volume = {288},
        pages = {205-220},
          doi = {10.1086/162783},
       adsurl = {https://ui.adsabs.harvard.edu/abs/1985ApJ...288..205A}
}

@ARTICLE{Baribaud1992,
       author = {{Baribaud}, T. and {Alloin}, D. and {Glass}, I. and {Pelat}, D.},
        title = "{Variability pattern from X-ray to IR wavelengths in the active nucleus ofNGC 1566.}",
      journal = {\aap},
         year = 1992,
        month = mar,
       volume = {256},
        pages = {375-398},
       adsurl = {https://ui.adsabs.harvard.edu/abs/1992A&A...256..375B}
}

@ARTICLE{Winkler1992,
       author = {{Winkler}, H.},
        title = "{Variability studies of Seyfert galaxies - II. Spectroscopy.}",
      journal = {\mnras},
         year = 1992,
        month = aug,
       volume = {257},
        pages = {677-688},
          doi = {10.1093/mnras/257.4.677},
       adsurl = {https://ui.adsabs.harvard.edu/abs/1992MNRAS.257..677W}
}

@ARTICLE{Ochmann2024,
       author = {{Ochmann}, M.~W. and {Kollatschny}, W. and {Probst}, M.~A. and {Romero-Colmenero}, E. and {Buckley}, D.~A.~H. and {Chelouche}, D. and {Chini}, R. and {Grupe}, D. and {Haas}, M. and {Kaspi}, S. and {Komossa}, S. and {Parker}, M.~L. and {Santos-Lleo}, M. and {Schartel}, N. and {Famula}, P.},
        title = "{The transient event in NGC 1566 from 2017 to 2019. I. An eccentric accretion disk and a turbulent, disk-dominated broad-line region unveiled by double-peaked Ca II and O I lines}",
      journal = {\aap},
         year = 2024,
        month = jun,
       volume = {686},
          eid = {A17},
        pages = {A17},
          doi = {10.1051/0004-6361/202348559},
archivePrefix = {arXiv},
       eprint = {2402.12054},
 primaryClass = {astro-ph.HE},
       adsurl = {https://ui.adsabs.harvard.edu/abs/2024A&A...686A..17O}
}

@ARTICLE{Quintana1975,
       author = {{Quintana}, H. and {Kaufmann}, P. and {Sasic}, J.~L.},
        title = "{Activity of the nucleus of the southern Seyfert galaxy NGC 1566}",
      journal = {\mnras},
         year = 1975,
        month = nov,
       volume = {173},
        pages = {57P-60P},
          doi = {10.1093/mnras/173.1.57P},
       adsurl = {https://ui.adsabs.harvard.edu/abs/1975MNRAS.173P..57Q}
}

@ARTICLE{Pastoriza1970,
       author = {{Pastoriza}, M. and {Gerola}, H.},
        title = "{Spectral Variation in the Seyfert Galaxy NGC 1566}",
      journal = {\aplett},
         year = 1970,
        month = jul,
       volume = {6},
        pages = {155},
       adsurl = {https://ui.adsabs.harvard.edu/abs/1970ApL.....6..155P}
}

@ARTICLE{Shobbrook1966,
       author = {{Shobbrook}, R.~R.},
        title = "{Southern groups and clusters of galaxies. I, Spectra and radial velocities of nineteen southern galaxies}",
      journal = {\mnras},
         year = 1966,
        month = jan,
       volume = {131},
        pages = {293},
          doi = {10.1093/mnras/131.2.293},
       adsurl = {https://ui.adsabs.harvard.edu/abs/1966MNRAS.131..293S}
}

@ARTICLE{Dodd2025,
       author = {{Dodd}, Sierra A. and {Huang}, Xiaoshan and {Davis}, Shane W. and {Ramirez-Ruiz}, Enrico},
        title = "{Perturbing AGN Accretion Disks with Stars and Moderately Massive Black Holes: Implications for Changing-Look AGN and Quasi-Periodic Eruptions}",
      journal = {arXiv e-prints},
         year = 2025,
        month = jun,
          eid = {arXiv:2506.19900},
        pages = {arXiv:2506.19900},
          doi = {10.48550/arXiv.2506.19900},
archivePrefix = {arXiv},
       eprint = {2506.19900},
 primaryClass = {astro-ph.HE},
       adsurl = {https://ui.adsabs.harvard.edu/abs/2025arXiv250619900D}
}

@ARTICLE{Evans2007,
       author = {{Evans}, P.~A. and {Beardmore}, A.~P. and {Page}, K.~L. and {Tyler}, L.~G. and {Osborne}, J.~P. and {Goad}, M.~R. and {O'Brien}, P.~T. and {Vetere}, L. and {Racusin}, J. and {Morris}, D. and {Burrows}, D.~N. and {Capalbi}, M. and {Perri}, M. and {Gehrels}, N. and {Romano}, P.},
        title = "{An online repository of Swift/XRT light curves of {\ensuremath{\gamma}}-ray bursts}",
      journal = {\aap},
         year = 2007,
        month = jul,
       volume = {469},
       number = {1},
        pages = {379-385},
          doi = {10.1051/0004-6361:20077530},
archivePrefix = {arXiv},
       eprint = {0704.0128},
 primaryClass = {astro-ph},
       adsurl = {https://ui.adsabs.harvard.edu/abs/2007A&A...469..379E}
}

@ARTICLE{daSilva2017,
       author = {{da Silva}, Patr{\'\i}cia and {Steiner}, J.~E. and {Menezes}, R.~B.},
        title = "{NGC 1566: analysis of the nuclear region from optical and near-infrared Integral Field Unit spectroscopy}",
      journal = {\mnras},
         year = 2017,
        month = oct,
       volume = {470},
       number = {4},
        pages = {3850-3876},
          doi = {10.1093/mnras/stx1458},
archivePrefix = {arXiv},
       eprint = {1707.02680},
 primaryClass = {astro-ph.GA},
       adsurl = {https://ui.adsabs.harvard.edu/abs/2017MNRAS.470.3850D}
}

@ARTICLE{Guo2025,
       author = {{Guo}, Wei-Jian and {Zou}, Hu and {Greenwell}, Claire L. and {Alexander}, David M. and {Fawcett}, Victoria A. and {Pan}, Zhiwei and {Siudek}, Ma{\l}gorzata and {Aguilar}, Jessica Nicole and {Ahlen}, Steven and {Brooks}, David and {Claybaugh}, Todd and {Dawson}, Kyle and {de la Macorra}, Axel and {Doel}, Peter and {Font-Ribera}, Andreu and {Gazta{\~n}aga}, Enrique and {Gontcho A Gontcho}, Satya and {Gutierrez}, Gaston and {Kehoe}, Robert and {Kisner}, Theodore and {Landriau}, Martin and {Le Guillou}, Laurent and {Manera}, Marc and {Meisner}, Aaron and {Miquel}, Ramon and {Moustakas}, John and {Prada}, Francisco and {Rossi}, Graziano and {Sanchez}, Eusebio and {Schubnell}, Michael and {Sprayberry}, David and {Sui}, Jipeng and {Tarl{\'e}}, Gregory and {Weaver}, Benjamin Alan and {Xiao}, Yun-Ao and {Zou}, Siwei},
        title = "{Changing-look Active Galactic Nuclei from the Dark Energy Spectroscopic Instrument. II. Statistical Properties from the First Data Release}",
      journal = {\apjs},
         year = 2025,
        month = may,
       volume = {278},
       number = {1},
          eid = {28},
        pages = {28},
          doi = {10.3847/1538-4365/adc124},
archivePrefix = {arXiv},
       eprint = {2408.00402},
 primaryClass = {astro-ph.GA},
       adsurl = {https://ui.adsabs.harvard.edu/abs/2025ApJS..278...28G}
}

@ARTICLE{Parker2019,
       author = {{Parker}, M.~L. and {Schartel}, N. and {Grupe}, D. and {Komossa}, S. and {Harrison}, F. and {Kollatschny}, W. and {Mikula}, R. and {Santos-Lle{\'o}}, M. and {Tom{\'a}s}, L.},
        title = "{X-ray spectra reveal the reawakening of the repeat changing-look AGN NGC 1566}",
      journal = {\mnras},
         year = 2019,
        month = feb,
       volume = {483},
       number = {1},
        pages = {L88-L92},
          doi = {10.1093/mnrasl/sly224},
archivePrefix = {arXiv},
       eprint = {1811.10289},
 primaryClass = {astro-ph.HE},
       adsurl = {https://ui.adsabs.harvard.edu/abs/2019MNRAS.483L..88P}
}

@ARTICLE{Hagen2025,
       author = {{Hagen}, Scott and {Done}, Chris and {Cackett}, Edward M. and {Partington}, Ethan R. and {Edelson}, Rick and {Lewin}, Collin and {Kara}, Erin and {Gelbord}, Jonathan},
        title = "{Untangling the complex nature of AGN variability in Fairall 9}",
      journal = {\mnras},
         year = 2025,
        month = nov,
       volume = {544},
       number = {1},
        pages = {1012-1037},
          doi = {10.1093/mnras/staf1751},
archivePrefix = {arXiv},
       eprint = {2509.25324},
 primaryClass = {astro-ph.HE},
       adsurl = {https://ui.adsabs.harvard.edu/abs/2025MNRAS.544.1012H}
}

@ARTICLE{Edelson2019,
       author = {{Edelson}, R. and {Gelbord}, J. and {Cackett}, E. and {Peterson}, B.~M. and {Horne}, K. and {Barth}, A.~J. and {Starkey}, D.~A. and {Bentz}, M. and {Brandt}, W.~N. and {Goad}, M. and {Joner}, M. and {Korista}, K. and {Netzer}, H. and {Page}, K. and {Uttley}, P. and {Vaughan}, S. and {Breeveld}, A. and {Cenko}, S.~B. and {Done}, C. and {Evans}, P. and {Fausnaugh}, M. and {Ferland}, G. and {Gonzalez-Buitrago}, D. and {Gropp}, J. and {Grupe}, D. and {Kaastra}, J. and {Kennea}, J. and {Kriss}, G. and {Mathur}, S. and {Mehdipour}, M. and {Mudd}, D. and {Nousek}, J. and {Schmidt}, T. and {Vestergaard}, M. and {Villforth}, C.},
        title = "{The First Swift Intensive AGN Accretion Disk Reverberation Mapping Survey}",
      journal = {\apj},
         year = 2019,
        month = jan,
       volume = {870},
       number = {2},
          eid = {123},
        pages = {123},
          doi = {10.3847/1538-4357/aaf3b4},
archivePrefix = {arXiv},
       eprint = {1811.07956},
 primaryClass = {astro-ph.HE},
       adsurl = {https://ui.adsabs.harvard.edu/abs/2019ApJ...870..123E}
}

@ARTICLE{Bentz2013,
       author = {{Bentz}, Misty C. and {Denney}, Kelly D. and {Grier}, Catherine J. and {Barth}, Aaron J. and {Peterson}, Bradley M. and {Vestergaard}, Marianne and {Bennert}, Vardha N. and {Canalizo}, Gabriela and {De Rosa}, Gisella and {Filippenko}, Alexei V. and {Gates}, Elinor L. and {Greene}, Jenny E. and {Li}, Weidong and {Malkan}, Matthew A. and {Pogge}, Richard W. and {Stern}, Daniel and {Treu}, Tommaso and {Woo}, Jong-Hak},
        title = "{The Low-luminosity End of the Radius-Luminosity Relationship for Active Galactic Nuclei}",
      journal = {\apj},
         year = 2013,
        month = apr,
       volume = {767},
       number = {2},
          eid = {149},
        pages = {149},
          doi = {10.1088/0004-637X/767/2/149},
archivePrefix = {arXiv},
       eprint = {1303.1742},
 primaryClass = {astro-ph.CO},
       adsurl = {https://ui.adsabs.harvard.edu/abs/2013ApJ...767..149B}
}

@ARTICLE{Komossa2026,
       author = {{Komossa}, S. and {Grupe}, D. and {Marziani}, P. and {Popovi{\'c}}, L. {\v{C}}. and {Mar{\v{c}}eta-Mandi{\'c}}, S. and {Bon}, E. and {Ili{\'c}}, D. and {Kova{\v{c}}evi{\'c}}, A.~B. and {Kraus}, A. and {Haiman}, Z. and {Petrecca}, V. and {De Cicco}, D. and {Dimitrijevi{\'c}}, M.~S. and {Sre{\'c}kovi{\'c}}, V.~A. and {Kova{\v{c}}evi{\'c} Doj{\v{c}}inovi{\'c}}, J. and {Pannikkote}, M. and {Bon}, N. and {Gupta}, K.~K. and {Iacob}, F.},
        title = "{The extremes of AGN variability: Outbursts, deep fades, changing looks, exceptional spectral states, and semi-periodicities}",
      journal = {Advances in Space Research},
         year = 2026,
        month = feb,
       volume = {77},
       number = {3},
        pages = {4041-4058},
          doi = {10.1016/j.asr.2025.04.058},
archivePrefix = {arXiv},
       eprint = {2408.00089},
 primaryClass = {astro-ph.HE},
       adsurl = {https://ui.adsabs.harvard.edu/abs/2026AdSpR..77.4041K}
}

@ARTICLE{Grupe2026,
       author = {{Grupe}, D. and {Komossa}, S. and {Zheng}, W. and {Filippenko}, A.~V. and {Brink}, T.~G. and {Schartel}, N. and {Wang}, J. and {Oknyansky}, V. and {Dodin}, A.~V. and {Wolsing}, S. and {Elien}, E. and {Tapper}, C. and {Lynam}, P.},
        title = "{In the Eye of the Storm: The Third Giant X-Ray Outburst of the Extreme Changing-look AGN IC 3599}",
      journal = {\apjl},
         year = 2026,
        month = jul,
       volume = {1006},
       number = {1},
          eid = {L23},
        pages = {L23},
          doi = {10.3847/2041-8213/ae884e},
archivePrefix = {arXiv},
       eprint = {2607.09036},
 primaryClass = {astro-ph.GA},
       adsurl = {https://ui.adsabs.harvard.edu/abs/2026ApJ..1006L..23G}
}

@ARTICLE{Hagen2023,
       author = {{Hagen}, Scott and {Done}, Chris},
        title = "{Modelling continuum reverberation in active galactic nuclei: a spectral-timing analysis of the ultraviolet variability through X-ray reverberation in Fairall 9}",
      journal = {\mnras},
         year = 2023,
        month = may,
       volume = {521},
       number = {1},
        pages = {251-268},
          doi = {10.1093/mnras/stad504},
archivePrefix = {arXiv},
       eprint = {2210.04924},
 primaryClass = {astro-ph.HE},
       adsurl = {https://ui.adsabs.harvard.edu/abs/2023MNRAS.521..251H}
}

@ARTICLE{Gunasekera2025,
       author = {{Gunasekera}, C.~M. and {van Hoof}, P.~A.~M. and {Dehghanian}, M. and {Chakraborty}, P. and {Shaw}, G. and {Bianchi}, S. and {Chatzikos}, M. and {Tsujimoto}, M. and {Ferland}, G.~J.},
        title = "{The 2025 release of Cloudy}",
      journal = {\rmxaa},
         year = 2025,
        month = nov,
       volume = {61},
        pages = {120-133},
          doi = {10.22201/ia.01851101p.2025.61.03.01},
archivePrefix = {arXiv},
       eprint = {2508.01102},
 primaryClass = {astro-ph.GA},
       adsurl = {https://ui.adsabs.harvard.edu/abs/2025RMxAA..61c.120G}
}

@ARTICLE{Bianchi2022,
       author = {{Bianchi}, Stefano and {Chiaberge}, Marco and {Laor}, Ari and {Antonucci}, Robert and {Bagul}, Atharva and {Capetti}, Alessandro},
        title = "{NGC 3147: a prototypical low-luminosity active galactic nucleus with double-peaked optical and ultraviolet lines}",
      journal = {\mnras},
         year = 2022,
        month = nov,
       volume = {516},
       number = {4},
        pages = {5775-5784},
          doi = {10.1093/mnras/stac2290},
archivePrefix = {arXiv},
       eprint = {2209.01807},
 primaryClass = {astro-ph.HE},
       adsurl = {https://ui.adsabs.harvard.edu/abs/2022MNRAS.516.5775B}
}

@ARTICLE{Mehdipour2011,
       author = {{Mehdipour}, M. and {Branduardi-Raymont}, G. and {Kaastra}, J.~S. and {Petrucci}, P.~O. and {Kriss}, G.~A. and {Ponti}, G. and {Blustin}, A.~J. and {Paltani}, S. and {Cappi}, M. and {Detmers}, R.~G. and {Steenbrugge}, K.~C.},
        title = "{Multiwavelength campaign on Mrk 509. IV. Optical-UV-X-ray variability and the nature of the soft X-ray excess}",
      journal = {\aap},
         year = 2011,
        month = oct,
       volume = {534},
          eid = {A39},
        pages = {A39},
          doi = {10.1051/0004-6361/201116875},
archivePrefix = {arXiv},
       eprint = {1107.0659},
 primaryClass = {astro-ph.CO},
       adsurl = {https://ui.adsabs.harvard.edu/abs/2011A&A...534A..39M}
}

@ARTICLE{Ho2008,
       author = {{Ho}, L.~C.},
        title = "{Nuclear activity in nearby galaxies.}",
      journal = {\araa},
         year = 2008,
        month = sep,
       volume = {46},
        pages = {475-539},
          doi = {10.1146/annurev.astro.45.051806.110546},
archivePrefix = {arXiv},
       eprint = {0803.2268},
 primaryClass = {astro-ph},
       adsurl = {https://ui.adsabs.harvard.edu/abs/2008ARA&A..46..475H}
}

@ARTICLE{Jana2026,
       author = {{Jana}, Arghajit and {Ricci}, Claudio and {Tortosa}, Alessia and {Dimopoulos}, George and {Trakhtenbrot}, Benny and {Bauer}, Franz E. and {Temple}, Matthew J. and {Koss}, Michael and {Gupta}, Kriti Kamal and {Chang}, Hsian-Kuang and {Diaz}, Yaherlyn and {Illic}, Dragana and {Kallov{\'a}}, Krist{\'\i}na and {Shablovinskaya}, Elena},
        title = "{Multiwavelength properties of changing-state active galactic nuclei: I. The evolution of soft excess and X-ray continuum}",
      journal = {\aap},
         year = 2026,
        month = mar,
       volume = {707},
          eid = {A213},
        pages = {A213},
          doi = {10.1051/0004-6361/202556654},
archivePrefix = {arXiv},
       eprint = {2601.07337},
 primaryClass = {astro-ph.HE},
       adsurl = {https://ui.adsabs.harvard.edu/abs/2026A&A...707A.213J}
}

@ARTICLE{Chen2025,
       author = {{Chen}, Shi-Jiang and {Buchner}, Johannes and {Liu}, Teng and {Hagen}, Scott and {Waddell}, Sophia G.~H. and {Nandra}, Kirpal and {Salvato}, Mara and {Igo}, Zsofi and {Aydar}, Catarina and {Merloni}, Andrea and {Ni}, Qingling and {Kang}, Jia-Lai and {Cai}, Zhen-Yi and {Wang}, Jun-Xian and {Li}, Ruancun and {Ramos-Ceja}, Miriam E. and {Sanders}, Jeremy and {Georgakakis}, Antonis and {Zhang}, Yi},
        title = "{The average soft X-ray spectra of eROSITA active galactic nuclei}",
      journal = {\aap},
         year = 2025,
        month = sep,
       volume = {701},
          eid = {A144},
        pages = {A144},
          doi = {10.1051/0004-6361/202554737},
archivePrefix = {arXiv},
       eprint = {2506.17150},
 primaryClass = {astro-ph.HE},
       adsurl = {https://ui.adsabs.harvard.edu/abs/2025A&A...701A.144C}
}

@ARTICLE{Hart2023,
       author = {{Hart}, K. and {Shappee}, B.~J. and {Hey}, D. and {Kochanek}, C.~S. and {Stanek}, K.~Z. and {Lim}, L. and {Dobbs}, S. and {Tucker}, M. and {Jayasinghe}, T. and {Beacom}, J.~F. and {Boright}, T. and {Holoien}, T. and {Ong}, J.~M. Joel and {Prieto}, J.~L. and {Thompson}, T.~A. and {Will}, D.},
        title = "{ASAS-SN Sky Patrol V2.0}",
      journal = {arXiv e-prints},
         year = 2023,
        month = apr,
          eid = {arXiv:2304.03791},
        pages = {arXiv:2304.03791},
          doi = {10.48550/arXiv.2304.03791},
archivePrefix = {arXiv},
       eprint = {2304.03791},
 primaryClass = {astro-ph.IM},
       adsurl = {https://ui.adsabs.harvard.edu/abs/2023arXiv230403791H}
}

@ARTICLE{Korista2019,
       author = {{Korista}, K.~T. and {Goad}, M.~R.},
        title = "{Quantifying the impact of variable BLR diffuse continuum contributions on measured continuum interband delays}",
      journal = {\mnras},
         year = 2019,
        month = nov,
       volume = {489},
       number = {4},
        pages = {5284-5300},
          doi = {10.1093/mnras/stz2330},
archivePrefix = {arXiv},
       eprint = {1908.07757},
 primaryClass = {astro-ph.GA},
       adsurl = {https://ui.adsabs.harvard.edu/abs/2019MNRAS.489.5284K}
}

@ARTICLE{Korista2001,
       author = {{Korista}, Kirk T. and {Goad}, Michael R.},
        title = "{The Variable Diffuse Continuum Emission of Broad-Line Clouds}",
      journal = {\apj},
         year = 2001,
        month = jun,
       volume = {553},
       number = {2},
        pages = {695-708},
          doi = {10.1086/320964},
archivePrefix = {arXiv},
       eprint = {astro-ph/0101117},
 primaryClass = {astro-ph},
       adsurl = {https://ui.adsabs.harvard.edu/abs/2001ApJ...553..695K}
}

@ARTICLE{Laor1989,
       author = {{Laor}, Ari and {Netzer}, Hagai},
        title = "{Massive thin accretion discs. - I. Calculated spectra.}",
      journal = {\mnras},
         year = 1989,
        month = jun,
       volume = {238},
        pages = {897-916},
          doi = {10.1093/mnras/238.3.897},
       adsurl = {https://ui.adsabs.harvard.edu/abs/1989MNRAS.238..897L}
}

@ARTICLE{Raj2021,
       author = {{Raj}, A. and {Nixon}, C.~J. and {Do{\u{g}}an}, S.},
        title = "{Disk Tearing: Numerical Investigation of Warped Disk Instability}",
      journal = {\apj},
         year = 2021,
        month = mar,
       volume = {909},
       number = {1},
          eid = {81},
        pages = {81},
          doi = {10.3847/1538-4357/abdc24},
archivePrefix = {arXiv},
       eprint = {2101.05824},
 primaryClass = {astro-ph.HE},
       adsurl = {https://ui.adsabs.harvard.edu/abs/2021ApJ...909...81R}
}

@ARTICLE{Petrucci2020,
       author = {{Petrucci}, P.-O. and {Gronkiewicz}, D. and {Rozanska}, A. and {Belmont}, R. and {Bianchi}, S. and {Czerny}, B. and {Matt}, G. and {Malzac}, J. and {Middei}, R. and {De Rosa}, A. and {Ursini}, F. and {Cappi}, M.},
        title = "{Radiation spectra of warm and optically thick coronae in AGNs}",
      journal = {\aap},
         year = 2020,
        month = feb,
       volume = {634},
          eid = {A85},
        pages = {A85},
          doi = {10.1051/0004-6361/201937011},
archivePrefix = {arXiv},
       eprint = {2001.02026},
 primaryClass = {astro-ph.HE},
       adsurl = {https://ui.adsabs.harvard.edu/abs/2020A&A...634A..85P}
}

@ARTICLE{Middei2023,
       author = {{Middei}, R. and {Petrucci}, P.-O. and {Bianchi}, S. and {Ursini}, F. and {Matzeu}, G.~A. and {Vagnetti}, F. and {Tortosa}, A. and {Marinucci}, A. and {Matt}, G. and {Piconcelli}, E. and {De Rosa}, A. and {De Marco}, B. and {Reeves}, J. and {Perri}, M. and {Guainazzi}, M. and {Cappi}, M. and {Done}, C.},
        title = "{Tracking the spectral properties across the different epochs in ESO 511-G030}",
      journal = {arXiv e-prints},
         year = 2023,
        month = feb,
          eid = {arXiv:2302.03705},
        pages = {arXiv:2302.03705},
          doi = {10.48550/arXiv.2302.03705},
archivePrefix = {arXiv},
       eprint = {2302.03705},
 primaryClass = {astro-ph.HE},
       adsurl = {https://ui.adsabs.harvard.edu/abs/2023arXiv230203705M}
}

@ARTICLE{Hagen2024b,
       author = {{Hagen}, Scott and {Done}, Chris and {Silverman}, John D. and {Li}, Junyao and {Liu}, Teng and {Ren}, Wenke and {Buchner}, Johannes and {Merloni}, Andrea and {Nagao}, Tohru and {Salvato}, Mara},
        title = "{Systematic collapse of the accretion disc across the supermassive black hole population}",
      journal = {\mnras},
         year = 2024,
        month = nov,
       volume = {534},
       number = {3},
        pages = {2803-2818},
          doi = {10.1093/mnras/stae2272},
archivePrefix = {arXiv},
       eprint = {2406.06674},
 primaryClass = {astro-ph.HE},
       adsurl = {https://ui.adsabs.harvard.edu/abs/2024MNRAS.534.2803H}
}

@ARTICLE{Poole2008,
       author = {{Poole}, T.~S. and {Breeveld}, A.~A. and {Page}, M.~J. and {Landsman}, W. and {Holland}, S.~T. and {Roming}, P. and {Kuin}, N.~P.~M. and {Brown}, P.~J. and {Gronwall}, C. and {Hunsberger}, S. and {Koch}, S. and {Mason}, K.~O. and {Schady}, P. and {vanden Berk}, D. and {Blustin}, A.~J. and {Boyd}, P. and {Broos}, P. and {Carter}, M. and {Chester}, M.~M. and {Cucchiara}, A. and {Hancock}, B. and {Huckle}, H. and {Immler}, S. and {Ivanushkina}, M. and {Kennedy}, T. and {Marshall}, F. and {Morgan}, A. and {Pandey}, S.~B. and {de Pasquale}, M. and {Smith}, P.~J. and {Still}, M.},
        title = "{Photometric calibration of the Swift ultraviolet/optical telescope}",
      journal = {\mnras},
         year = 2008,
        month = jan,
       volume = {383},
       number = {2},
        pages = {627-645},
          doi = {10.1111/j.1365-2966.2007.12563.x},
archivePrefix = {arXiv},
       eprint = {0708.2259},
 primaryClass = {astro-ph},
       adsurl = {https://ui.adsabs.harvard.edu/abs/2008MNRAS.383..627P}
}

@ARTICLE{Cackett2007,
       author = {{Cackett}, Edward M. and {Horne}, Keith and {Winkler}, Hartmut},
        title = "{Testing thermal reprocessing in active galactic nuclei accretion discs}",
      journal = {\mnras},
         year = 2007,
        month = sep,
       volume = {380},
       number = {2},
        pages = {669-682},
          doi = {10.1111/j.1365-2966.2007.12098.x},
archivePrefix = {arXiv},
       eprint = {0706.1464},
 primaryClass = {astro-ph},
       adsurl = {https://ui.adsabs.harvard.edu/abs/2007MNRAS.380..669C}
}

@ARTICLE{HI4PI2016,
       author = {{HI4PI Collaboration} and {Ben Bekhti}, N. and {Fl{\"o}er}, L. and {Keller}, R. and {Kerp}, J. and {Lenz}, D. and {Winkel}, B. and {Bailin}, J. and {Calabretta}, M.~R. and {Dedes}, L. and {Ford}, H.~A. and {Gibson}, B.~K. and {Haud}, U. and {Janowiecki}, S. and {Kalberla}, P.~M.~W. and {Lockman}, F.~J. and {McClure-Griffiths}, N.~M. and {Murphy}, T. and {Nakanishi}, H. and {Pisano}, D.~J. and {Staveley-Smith}, L.},
        title = "{HI4PI: A full-sky H I survey based on EBHIS and GASS}",
      journal = {\aap},
         year = 2016,
        month = oct,
       volume = {594},
          eid = {A116},
        pages = {A116},
          doi = {10.1051/0004-6361/201629178},
archivePrefix = {arXiv},
       eprint = {1610.06175},
 primaryClass = {astro-ph.GA},
       adsurl = {https://ui.adsabs.harvard.edu/abs/2016A&A...594A.116H}
}

@ARTICLE{Drake2009,
       author = {{Drake}, A.~J. and {Djorgovski}, S.~G. and {Mahabal}, A. and {Beshore}, E. and {Larson}, S. and {Graham}, M.~J. and {Williams}, R. and {Christensen}, E. and {Catelan}, M. and {Boattini}, A. and {Gibbs}, A. and {Hill}, R. and {Kowalski}, R.},
        title = "{First Results from the Catalina Real-Time Transient Survey}",
      journal = {\apj},
         year = 2009,
        month = may,
       volume = {696},
       number = {1},
        pages = {870-884},
          doi = {10.1088/0004-637X/696/1/870},
archivePrefix = {arXiv},
       eprint = {0809.1394},
 primaryClass = {astro-ph},
       adsurl = {https://ui.adsabs.harvard.edu/abs/2009ApJ...696..870D}
}

\appendix
\section{Host galaxy subtraction}
In Section \ref{sec:host} we subtract the host galaxy in multiple ways. We first tried a two-dimensional decomposition of high resolution HST UV imaging using \texttt{GALFIT} \citep{Peng2002}. The fit was applied to the HST/WFC3 F275W image over the central $12^{\prime\prime}\times 12^{\prime\prime}$ region. The model consisted of a PSF component representing the unresolved nuclear emission, a Sérsic component representing the extended host-galaxy emission, and a constant sky background. The model was convolved with the corresponding F275W PSF.

Figure~\ref{fig:galfit_result} shows the observed image, the best-fitting model, and the residual image. Significant residual structures remain in the vicinity of the nucleus, demonstrating that a simple PSF+Sérsic decomposition does not adequately describe the complex and non-axisymmetric morphology of the central region. We therefore do not use the quantitative \texttt{GALFIT} results in the subsequent host-galaxy subtraction.

As an alternative to the image decomposition, we also examined a linear variability-based flux–flux decomposition. 
The residuals from these linear relations are shown for all six UVOT bands in Fig.~\ref{fig:variability_residual}. Clear systematic deviations are present at the faint end. In the UVW2 and UVM2 bands, the linear relation tends to underestimate the observed flux at the lowest flux levels, whereas in UVW1, U, B, and V it tends to overestimate the flux. The deviations are particularly evident in the optical bands, where the inferred host contribution can become sufficiently large that subtracting it produces negative AGN fluxes at the faintest epochs.

The coherent, flux-dependent residual patterns across the six bands indicate that the UV/optical variability cannot be described by a single component with a fixed spectral shape. Instead, the shape of the variable component changes with source luminosity. We therefore only use the linear flux–flux decomposition as a systematic uncertainty in the host-galaxy contribution, which is shown in Table \ref{tab:faintest5_and_host}.

\section{Long-term behaviour of NGC1566}

From top to bottom, the panels show the spectroscopically inferred broad-line state\citep{Shobbrook1966, Pastoriza1970, Quintana1975, Alloin1985, Winkler1992, Ochmann2024}, the historical $B$-band photometry compiled by \citet{Quintana1975}, the historical $JHKL$ photometry presented by \citet{Baribaud1992}, the 3700 and 5000~\AA\ continuum flux densities and broad H$\alpha$ and H$\beta$ line fluxes (scaled by a factor of 3) from \citet{Alloin1985}, the CRTS unfiltered light curve, the ASAS-SN $V$- and $g$-band light curves, and the Swift/UVOT UVW2 flux-density light curve.

\renewcommand{\thefigure}{A\arabic{figure}}
\renewcommand{\thetable}{A\arabic{table}}
\setcounter{figure}{0}
\setcounter{table}{0}

\begin{table}
\centering
\caption{Extraction-region properties as a function of the observed \textit{Swift}/XRT count rate in PC mode.}
\label{tab:xrt_regions}
\begin{tabular}{lcc}
\hline
Region shape & Radius (inner radius) & Count rate \\
             & pixel                 & counts s$^{-1}$ \\
\hline
Circle  & 5        & $<0.0005$ \\
Circle  & 7        & $0.0005$--$0.001$ \\
Circle  & 9        & $0.001$--$0.005$ \\
Circle  & 12       & $0.005$--$0.01$ \\
Circle  & 15       & $0.01$--$0.05$ \\
Circle  & 20       & $0.05$--$0.1$ \\
Circle  & 25       & $0.1$--$0.6$ \\
Annulus & 25 (2)   & $0.6$--$1.4$ \\
Annulus & 25 (3)   & $1.4$--$1.7$ \\
Annulus & 25 (4)   & $1.7$--$2.8$ \\
Annulus & 25 (5)   & $2.8$--$3.2$ \\
Annulus & 25 (6)   & $>3.2$ \\
\hline
\end{tabular}

\vspace{0.5ex}
\begin{minipage}{0.9\linewidth}
\footnotesize
Note. For annular regions, the first value gives the outer radius, and the value in parentheses gives the excluded inner radius. 
\end{minipage}
\end{table}

\begin{table*}
\centering
\caption{Host galaxy estimation}
\label{tab:faintest5_and_host}

\begin{tabular}{cccccccc}
\hline
\hline
OBSID &
MJD &
UVW2 &
UVM2 &
UVW1 &
U &
B &
V \\
&
&
\multicolumn{6}{c}{
$F_{\lambda}\,[10^{-15}\,\mathrm{erg\,cm^{-2}\,s^{-1}\,\AA^{-1}}]$
}
\\
\hline

00033411002 &
56912.403 &
$1.601 \pm 0.622$ &
$3.563 \pm 0.125$ &
$2.893 \pm 0.489$ &
$5.035 \pm 0.413$ &
$8.112 \pm 0.551$ &
$16.339 \pm 0.980$ \\

00033411021 &
56940.643 &
$1.670 \pm 0.185$ &
$1.609 \pm 0.086$ &
$3.190 \pm 0.496$ &
$6.039 \pm 0.437$ &
$9.800 \pm 0.586$ &
$14.546 \pm 0.910$ \\

00033411013 &
56919.924 &
$1.870 \pm 0.515$ &
$1.802 \pm 0.100$ &
$3.317 \pm 0.502$ &
$5.088 \pm 0.403$ &
$9.130 \pm 0.560$ &
$15.284 \pm 0.937$ \\

00045604050 &
60236.440 &
$1.880 \pm 0.110$ &
$1.877 \pm 0.134$ &
$3.214 \pm 0.147$ &
$4.858 \pm 0.175$ &
$8.209 \pm 0.242$ &
$14.524 \pm 0.441$ \\

00045604069 &
61029.479 &
$1.883 \pm 0.104$ &
$1.568 \pm 0.131$ &
$3.043 \pm 0.137$ &
$5.010 \pm 0.171$ &
$10.785 \pm 0.288$ &
$14.705 \pm 0.429$ \\

\hline

\multicolumn{2}{c}{Host galaxy (faintest five)} &

$1.781 \pm 0.135$ &

$2.084 \pm 0.837$ &

$3.131 \pm 0.166$ &

$5.206 \pm 0.474$ &

$9.207 \pm 1.123$ &

$15.080 \pm 0.770$ \\

\multicolumn{2}{c}{Host galaxy (variability)} &

$0.165 \pm 1.616$ &

$0.436 \pm 1.648$ &

$2.511 \pm 0.620$ &

$5.190 \pm 0.016$ &

$10.800 \pm 1.590$ &

$16.085 \pm 1.010$ \\

\multicolumn{2}{c}{Host galaxy (total)} &

$1.781 \pm 2.289$ &

$2.084 \pm 2.477$ &

$3.131 \pm 0.893$ &

$5.206 \pm 0.474$ &

$9.207 \pm 2.517$ &

$15.080 \pm 1.620$ \\

\hline
\end{tabular}

\end{table*}











\begin{table*}
\centering
\caption{SED fitting results}
\label{tab: sed_fitting}
\renewcommand{\arraystretch}{1.3}
\begin{tabular}{lccccccc}
\hline
\hline
Spectrum &
MJD &
$\log \dot{m}$ &
$kT_{\rm e,warm}$ (keV) &
$\Gamma_{\rm hot}$ &
$\Gamma_{\rm warm}$ &
$R_{\rm hot}$ ($R_{\rm g}$) &
$R_{\rm warm}$ ($R_{\rm g}$) \\
\hline

00035880013 &
58330.9 &
$-1.155^{+0.016}_{-0.018}$ &
$0.313^{+0.096}_{-0.060}$ &
$1.450^{*}$ &
$2.578^{+0.036}_{-0.036}$ &
$68.2^{+4.7}_{-4.7}$ &
$448.1^{+35.1}_{-34.7}$ \\

00035880021 &
58346.4 &
$-1.180^{+0.016}_{-0.017}$ &
$0.100^{+0.019}_{-0.016}$ &
$1.500^{*}$ &
$2.185^{+0.065}_{-0.073}$ &
$96.3^{+8.1}_{-8.2}$ &
$487.4^{+487.2}_{-32.1}$ \\

00035880030 &
58357.5 &
$-1.556^{+0.093}_{-0.093}$ &
$0.114^{+0.040}_{-0.031}$ &
$1.585$ &
$2.329^{+0.151}_{-2.329}$ &
$74.8^{+24.2}_{-12.6}$ &
$208.4^{+91.5}_{-73.8}$ \\

00035880035 &
58362.1 &
$-1.685^{+0.074}_{-0.068}$ &
$0.220^{*}$ &
$1.432^{+0.039}_{-0.038}$ &
$2.400^{*}$ &
$108.8^{+18.9}_{-16.0}$ &
$162.9^{+37.2}_{-29.6}$ \\

00035880044 &
58380.0 &
$-1.796^{+0.139}_{-0.113}$ &
$0.220^{*}$ &
$1.495^{+0.071}_{-0.073}$ &
$2.400^{*}$ &
$117.7^{+30.7}_{-20.8}$ &
$210.0^{+87.1}_{-55.5}$ \\

00035880053 &
58395.6 &
$-1.918^{+0.127}_{-0.094}$ &
$0.220^{*}$ &
$1.544^{+0.064}_{-0.071}$ &
$2.557^{+0.355}_{-2.556}$ &
$107.1^{+34.3}_{-24.0}$ &
$165.7^{+103.3}_{-52.5}$ \\

bin\_004 &
58421.9 &
$-2.091^{+0.076}_{-0.066}$ &
$0.200^{*}$ &
$1.617^{+0.054}_{-0.053}$ &
$2.400^{*}$ &
$99.8^{+13.0}_{-11.3}$ &
$121.4^{+30.5}_{-121.4}$ \\

00035880072 &
58449.3 &
$-1.479^{+0.032}_{-0.152}$ &
$0.151^{+0.063}_{-0.042}$ &
$1.394^{+0.075}_{-0.013}$ &
$2.378^{+0.084}_{-0.092}$ &
$171.5^{+16.4}_{-49.5}$ &
$497.3^{+500.0}_{-150.5}$ \\

00035880076 &
58464.9 &
$-1.651^{+0.097}_{-0.083}$ &
$0.220^{*}$ &
$1.613^{+0.065}_{-0.068}$ &
$2.525^{+0.138}_{-0.228}$ &
$91.3^{+17.7}_{-13.8}$ &
$225.4^{+85.2}_{-70.3}$ \\

bin\_007 &
58486.8 &
$-2.030^{+0.053}_{-0.053}$ &
$0.220^{*}$ &
$1.610^{+0.036}_{-0.036}$ &
$2.400^{*}$ &
$153.0^{+17.0}_{-17.0}$ &
$=R_{\rm hot}$\\

bin\_009 &
58518.7 &
$-2.277^{+0.088}_{-0.088}$ &
$0.220^{*}$ &
$1.568^{+0.055}_{-0.055}$ &
$2.400^{*}$ &
$116.3^{+22.5}_{-22.5}$ &
$=R_{\rm hot}$ \\

bin\_010 &
58531.1 &
$-2.108^{+0.120}_{-0.120}$ &
$0.220^{*}$ &
$1.491^{+0.060}_{-0.060}$ &
$2.400^{*}$ &
$173.0^{+39.7}_{-39.7}$ &
$=R_{\rm hot}$\\

bin\_013 &
58576.3 &
$-2.364^{+0.096}_{-0.096}$ &
$0.220^{*}$ &
$1.562^{+0.058}_{-0.058}$ &
$2.400^{*}$ &
$129.0^{+24.8}_{-24.8}$ &
$136.7^{+109.2}_{-136.8}$ \\

00035880117 &
58631.9 &
$-1.528^{+0.070}_{-0.108}$ &
$0.220^{*}$ &
$1.475^{+0.057}_{-0.036}$ &
$2.473^{+1.036}_{-0.206}$ &
$295.6^{+59.2}_{-51.9}$ &
$500.0^{+500.0}_{-86.2}$ \\

bin\_015 &
58637.1 &
$-2.208^{+0.044}_{-0.044}$ &
$0.220^{*}$ &
$1.710^{+0.038}_{-0.038}$ &
$3.170^{+1.000}_{-1.000}$ &
$166.3^{+13.3}_{-13.3}$ &
$499.8^{+500.0}_{-117.0}$ \\

bin\_017 &
58652.5 &
$-2.555^{+0.110}_{-0.110}$ &
$0.220^{*}$ &
$1.709^{+0.102}_{-0.102}$ &
$2.400^{*}$ &
$115.3^{+24.2}_{-24.2}$ &
$247.8^{+186.5}_{-117.8}$ \\

00035880142 &
58681.3 &
$-1.693^{+0.092}_{-0.111}$ &
$0.220^{*}$ &
$1.509^{+0.064}_{-0.049}$ &
$2.359^{+1.601}_{-0.292}$ &
$330.5^{+74.4}_{-61.6}$ &
$500.0^{+500.0}_{-65.1}$ \\

00088910003 &
58716.6 &
$-2.162^{+0.124}_{-0.110}$ &
$0.220^{*}$ &
$1.525^{+0.062}_{-0.061}$ &
$2.400^{*}$ &
$195.2^{+47.7}_{-36.9}$ &
$=R_{\rm hot}$\\

bin\_020 &
58728.3 &
$-1.624^{+0.209}_{-0.209}$ &
$0.220^{*}$ &
$1.447^{+0.097}_{-0.097}$ &
$2.400^{*}$ &
$333.4^{+76.3}_{-76.3}$ &
$500.0^{+298.1}_{-298.1}$ \\

bin\_023 &
58849.9 &
$-2.122^{+0.072}_{-0.072}$ &
$0.220^{*}$ &
$1.556^{+0.041}_{-0.041}$ &
$2.400^{*}$ &
$219.6^{+26.3}_{-26.3}$ &
$242.6^{+90.5}_{-242.6}$ \\

00045604047 &
59031.2 &
$-1.737^{+0.138}_{-0.175}$ &
$0.220^{*}$ &
$1.425^{+0.082}_{-0.056}$ &
$2.400^{*}$ &
$331.1^{+70.3}_{-58.1}$ &
$405.0^{+405.4}_{-405.4}$ \\

\hline

bin\_029 (low) &
58558.5 &
$-2.229^{+0.010}_{-0.010}$ &
$0.220^{*}$ &
$1.500^{*}$ &
$2.400^{*}$ &
$128.9^{+15.5}_{-12.7}$ &
$163.5^{+24.0}_{-18.3}$ \\

bin\_030 (late low) &
58867.2 &
$-2.057^{+0.007}_{-0.007}$ &
$0.220^{*}$ &
$1.500^{*}$ &
$2.400^{*}$ &
$198.5^{+21.2}_{-17.1}$ &
$277.2^{+43.0}_{-30.9}$ \\

\hline
\multicolumn{6}{l}{\footnotesize $^{*}$Parameter held fixed and not fitted.} \\
\multicolumn{6}{l}{\footnotesize $\rm bin\_004: 00035880062,~ 00035880063,~ 00035880064,~ 00035880065$} \\
\multicolumn{6}{l}{\footnotesize $\rm bin\_007: 00035880081,~ 00035880083,~ 00035880084,~ 00035880085$} \\
\multicolumn{6}{l}{\footnotesize $\rm bin\_009: 00035880091,~ 00035880092,~ 00035880094,~ 00035880095,~ 00035880096$} \\
\multicolumn{6}{l}{\footnotesize $\rm bin\_010: 00035880097,~ 00035880098,~ 00035880099$} \\
\multicolumn{6}{l}{\footnotesize $\rm bin\_013: 00035880107,~ 00035880108,~ 00035880109,~ 00035880110,~ 00035880111$} \\
\multicolumn{6}{l}{\footnotesize $\rm bin\_015: 00035880120,~ 00035880121,~ 00035880122,~ 00035880126,~ 00035880125$} \\
\multicolumn{6}{l}{\footnotesize $\rm bin\_017: 00035880137,~ 00035880136,~ 00035880138$} \\
\multicolumn{6}{l}{\footnotesize $\rm bin\_020: 00045604010,~ 00045604011$} \\
\multicolumn{6}{l}{\footnotesize $\rm bin\_023: 00045604025,~ 00045604026,~ 00045604027,~ 00045604028,~ 00045604031$} \\
\multicolumn{8}{l}{\footnotesize $\rm bin\_029: 00035880091,~ 00035880092,~ 00035880094,~ 00035880095,~ 00035880096,~00035880097,~ 00035880098$} \\
\multicolumn{8}{l}{\footnotesize $\rm ~~~~~~~~~~~~~~~~00035880099,~ 00035880100,~ 00035880101,~00035880102,~ 00035880103,~ 00035880104,~ 00035880105$} \\
\multicolumn{8}{l}{\footnotesize $\rm ~~~~~~~~~~~~~~~~00035880106,~ 00035880107,~ 00035880108,~ 00035880109,~ 00035880110,~00035880111,~ 00035880112$} \\
\multicolumn{8}{l}{\footnotesize $\rm ~~~~~~~~~~~~~~~~00035880113,~ 00035880114,~ 00035880115,~ 00035880116$} \\
\multicolumn{8}{l}{\footnotesize $\rm bin\_030: 00035880133,~ 00035880131,~ 00035880134,~ 00035880137,~ 00035880136,~ 00035880138,~ 00035880139$} \\
\multicolumn{8}{l}{\footnotesize $\rm ~~~~~~~~~~~~~~~~00045604010,~ 00045604011,~ 00045604012,~ 00045604013,~ 00045604014,~ 00045604015,~ 00045604016$} \\
\multicolumn{8}{l}{\footnotesize $\rm ~~~~~~~~~~~~~~~~00045604017,~ 00045604020,~ 00045604021,~ 00045604022,~ 00045604023,~ 00045604024,~ 00045604025$} \\
\multicolumn{8}{l}{\footnotesize $\rm ~~~~~~~~~~~~~~~~00045604026,~ 00045604027,~ 00045604028,~ 00045604031,~ 00045604032,~ 00045604033,~ 00045604035$} \\
\multicolumn{8}{l}{\footnotesize $\rm ~~~~~~~~~~~~~~~~00045604036,~ 00045604037,~ 00045604038,~ 00045604040,~ 00045604039,~ 00045604041,~ 00045604042$} \\
\multicolumn{8}{l}{\footnotesize $\rm ~~~~~~~~~~~~~~~~00045604043,~ 00045604044,~ 00045604045,~ 00031742005,~ 00045604046,~ 00031742006,~ 00031742007$} \\
\multicolumn{8}{l}{\footnotesize $\rm ~~~~~~~~~~~~~~~~00045604049,~ 00031742008,~ 00031742009,~ 00031742011,~ 00031742012,~ 00031742013,~ 00031742015$} \\
\end{tabular}
\end{table*}

\begin{figure*}
    \centering
    \includegraphics[width=0.85\textwidth]{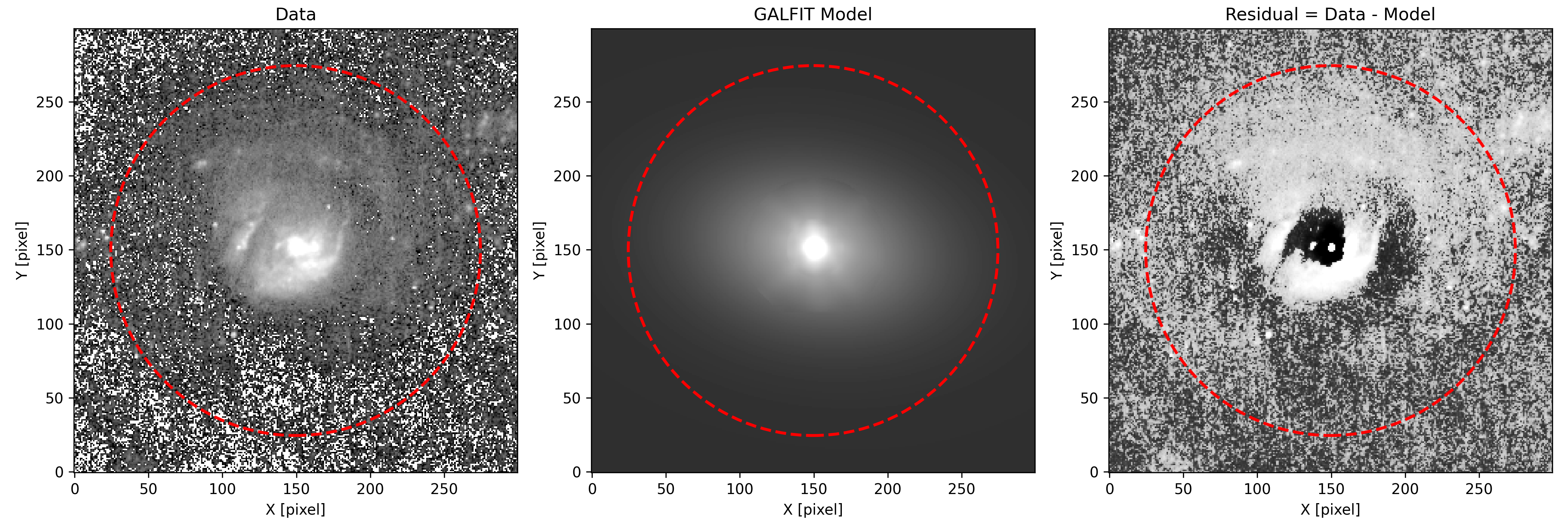}
    \caption{
    Two-dimensional image 
    decomposition of the central
    (12''$\times$12'') region
    of the HST image of NGC~1566 performed with
    \texttt{GALFIT}. The panels show 
    the observed image, the best-
    fitting model, and the residual
    image, respectively. The red
    dashed circle marks the (5'')
    -radius aperture used for the
    photometric extraction.}
    \label{fig:galfit_result}
\end{figure*}

\begin{figure*}
    \centering
    \includegraphics[width=0.8\textwidth]{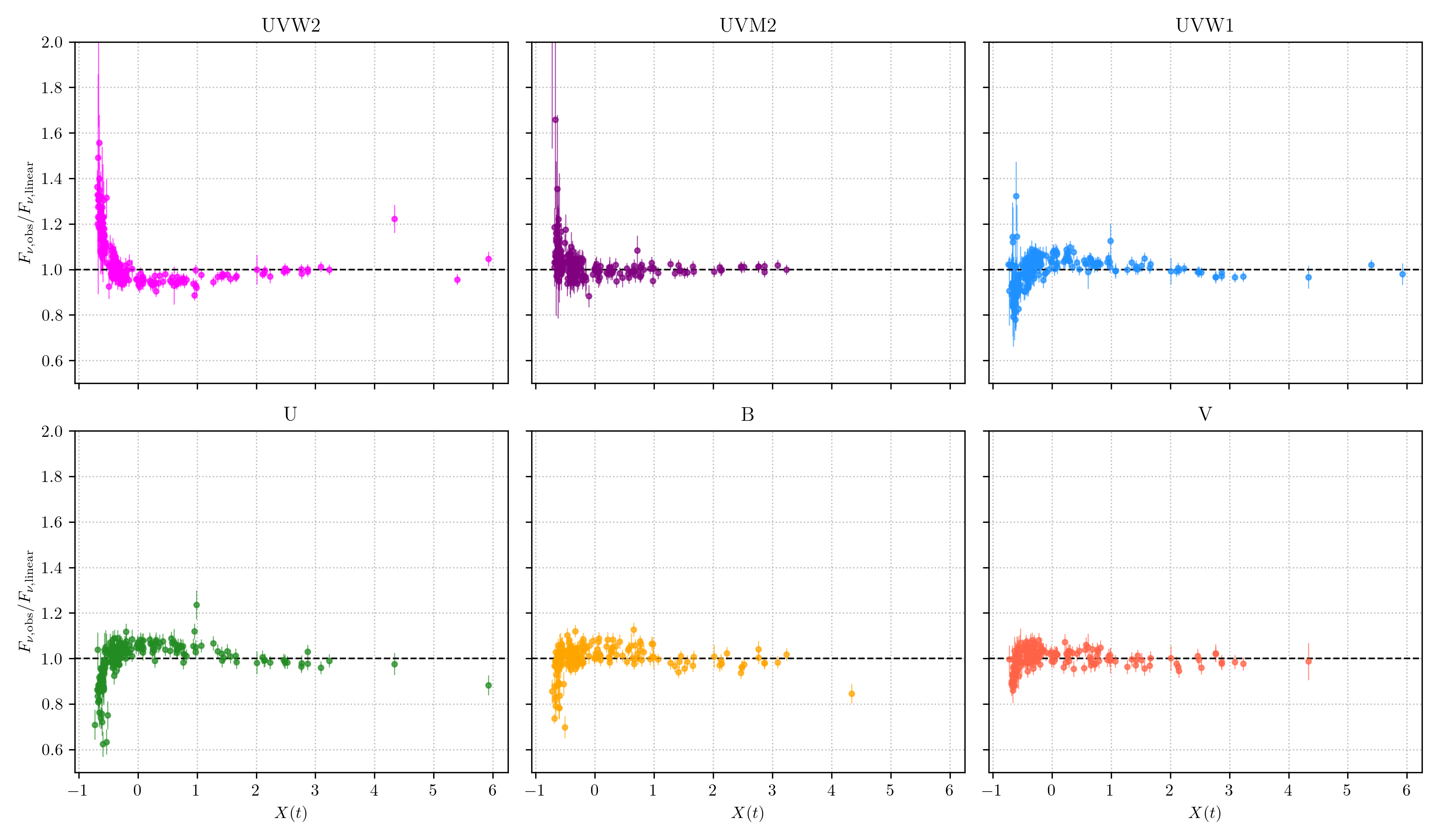}
    \caption{
    Residuals of the linear variability-based decomposition for the six UVOT bands. 
    Each panel shows the ratio between the observed flux density and the flux predicted by the best-fitting linear relation, 
    \(F_{\nu,{\rm obs}}/F_{\nu,{\rm linear}}\), as a function of the normalized light-curve shape \(X(t)\). 
    The horizontal dashed line marks perfect agreement between the observations and the linear model. 
    }
    \label{fig:variability_residual}
\end{figure*}

\begin{figure*}
    \centering
    \includegraphics[width=\textwidth]{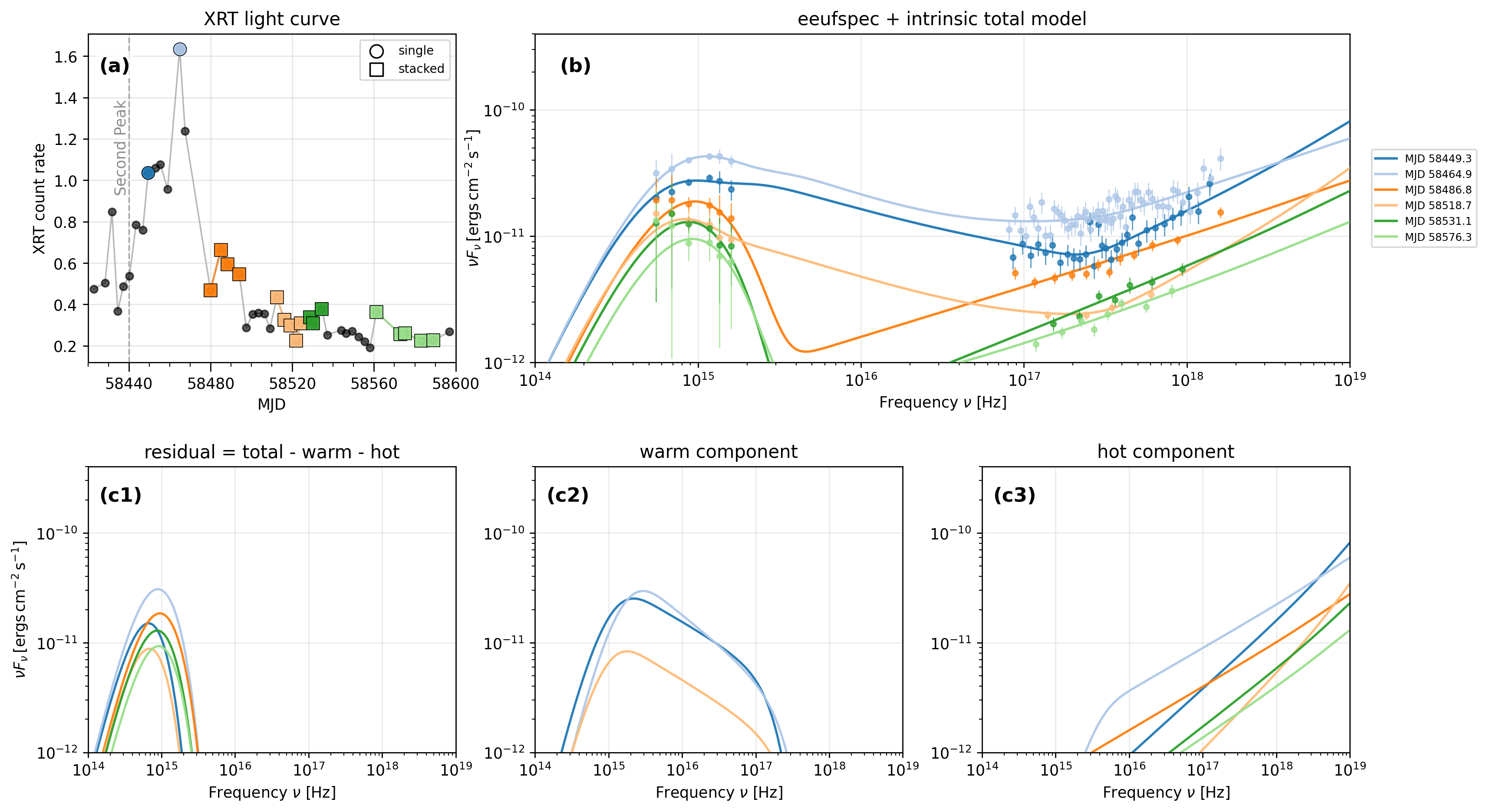}
    \caption{
    Evolution of the XRT light curve and broadband SED components of NGC~1566 between MJD 58400-58600. Panel (a) shows the Swift/XRT light curve, with the epochs used for SED fitting highlighted in colour. 
    Panel (b) presents the corresponding observed UVOT/XRT SEDs and best-fitting total \texttt{AGNSED} models. 
    Panels (c1)--(c3) decompose the observed model into the cold, warm, hot components. 
    }
    \label{fig:lightcurve_totalSED_components_2nd}
\end{figure*}

\begin{figure*}
    \centering
    \includegraphics[width=\textwidth]{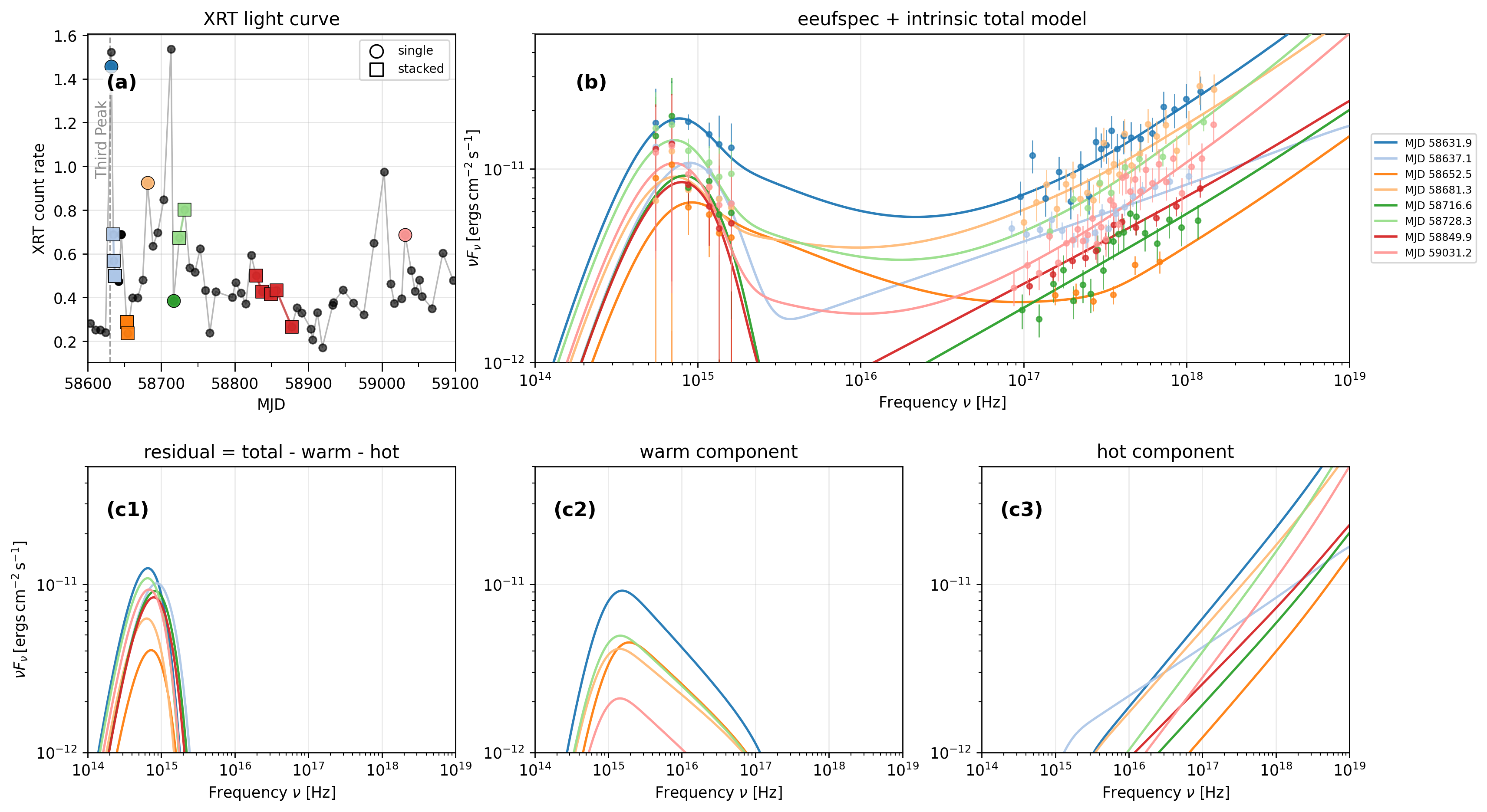}
    \caption{
    Evolution of the XRT light curve and broadband SED components of NGC~1566 between MJD 58600-59100. Panel (a) shows the Swift/XRT light curve, with the epochs used for SED fitting highlighted in colour. 
    Panel (b) presents the corresponding observed UVOT/XRT SEDs and best-fitting total \texttt{AGNSED} models. 
    Panels (c1)--(c3) decompose the observed model into the cold, warm, hot components. 
    }
    \label{fig:lightcurve_totalSED_components_3rd}
\end{figure*}

\begin{figure}
    \centering
    \includegraphics[width=\linewidth]{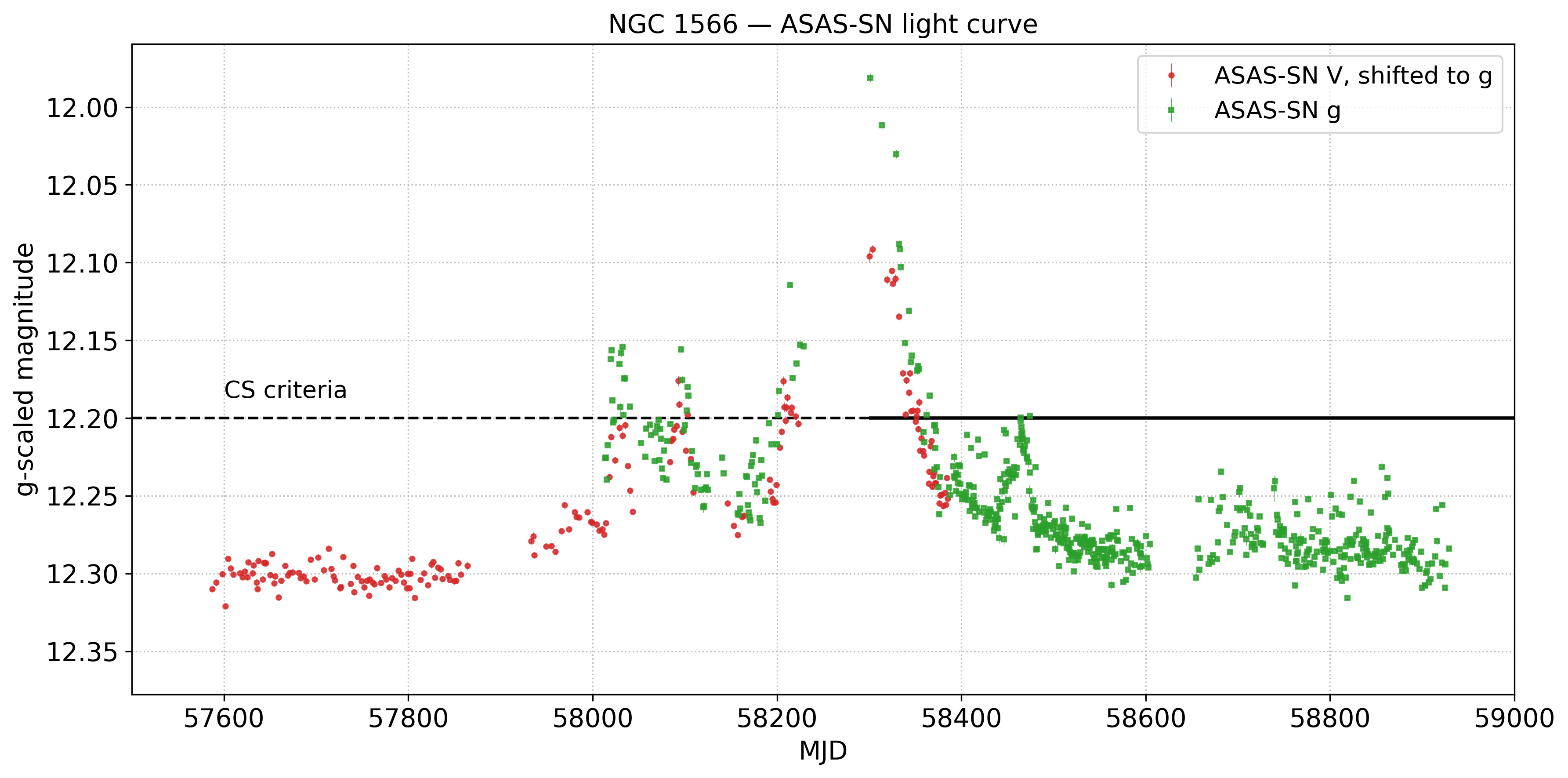}
    \caption{
    ASAS-SN optical light curve of NGC~1566.
    The $V$-band magnitudes are shifted by a constant offset to match the $g$-band magnitude scale, using the median magnitude difference (0.349 mag) over the overlapping time interval.
    The shifted $V$-band and original $g$-band measurements are shown in red and green, respectively. The horizontal line at $g=12.2$~mag marks the criteria where the changing state is observed (first and second peaks). The solid line covers the period with intensive \textit{Swift} monitoring, while the dashed line extends this to earlier times. This predicts there were other CS transitions at MJD~58000-58100 which were missed.}
    \label{fig:asassn_lc}
\end{figure}

\begin{figure*}
    \centering
    \includegraphics[width=\textwidth]{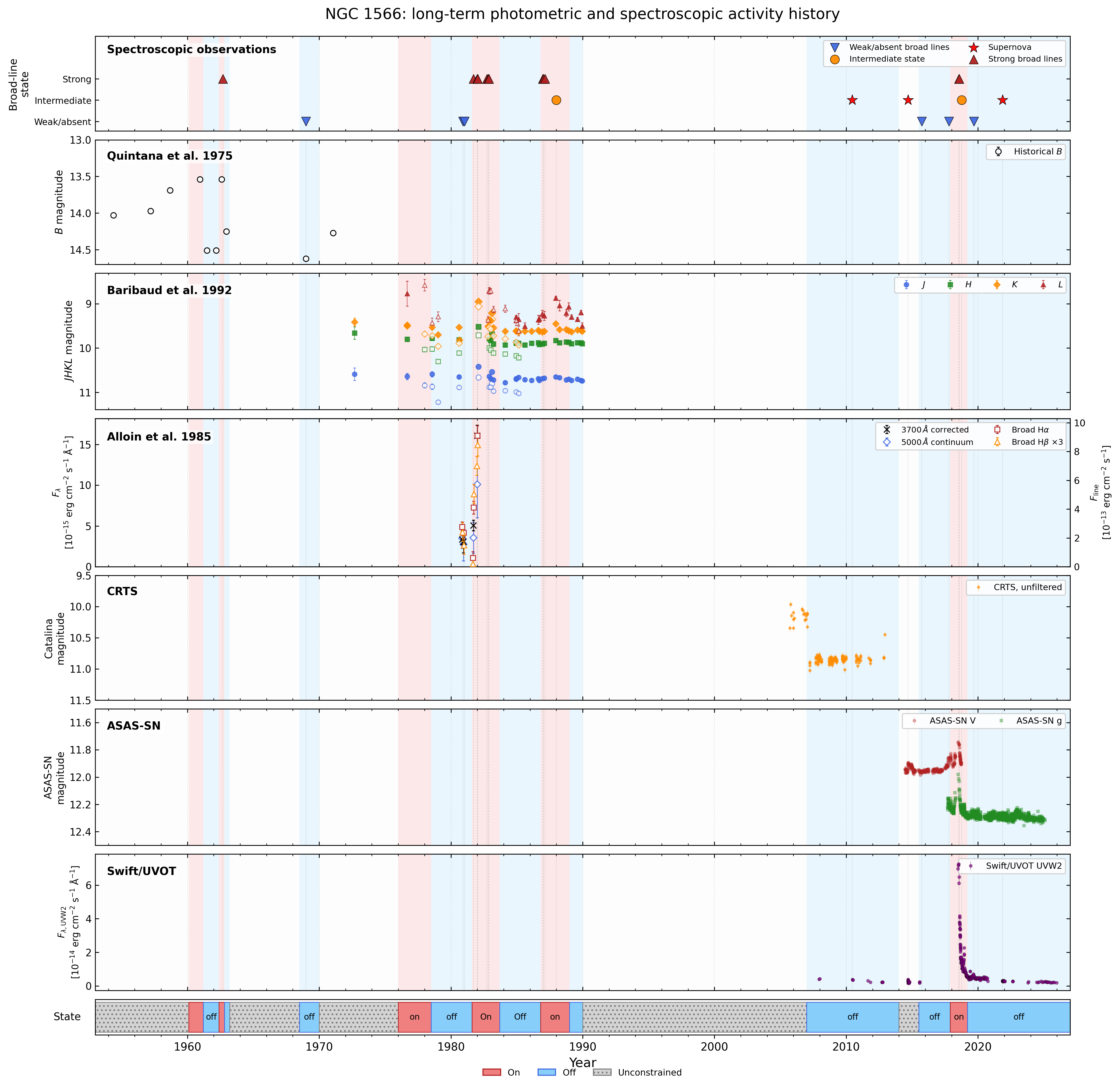}
    \caption{
    Long-term photometric and spectroscopic activity history of NGC~1566.
    }
    \label{fig:ngc1566_longterm_history}
\end{figure*}

\label{lastpage}
\end{document}